\documentclass[twocolumn,english,superscriptaddress]{revtex4-2}
\usepackage{lmodern}
\usepackage{lmodern}

\usepackage[T1]{fontenc}
\usepackage[latin9]{inputenc}
\usepackage{babel}
\usepackage{float}
\usepackage{amsmath}
\usepackage{amssymb}
\usepackage{graphicx}
\usepackage[unicode=true,pdfusetitle,
 bookmarks=true,bookmarksnumbered=false,bookmarksopen=false,
 breaklinks=false,pdfborder={0 0 1},backref=false,colorlinks=false]
 {hyperref}

\makeatletter

\providecommand{\tabularnewline}{\\}
\newcommand{\lyxdot}{.}

\usepackage{babel}

\makeatother

\begin{document}
\title{Self-similar solutions for the non-equilibrium nonlinear supersonic
Marshak wave problem}
\author{Menahem Krief}
\email{menahem.krief@mail.huji.ac.il}

\affiliation{Racah Institute of Physics, The Hebrew University, 9190401 Jerusalem,
Israel}
\affiliation{Department of Aerospace and Mechanical Engineering, University of
Notre Dame, Fitzpatrick Hall, Notre Dame, IN 46556, USA}
\author{Ryan G. McClarren}
\affiliation{Department of Aerospace and Mechanical Engineering, University of
Notre Dame, Fitzpatrick Hall, Notre Dame, IN 46556, USA}
\begin{abstract}
Similarity solutions to the nonlinear non-equilibrium Marshak wave
problem with a time dependent radiation driving source are presented.
The radiation transfer model used is the gray, non-equilibrium diffusion
approximation in the supersonic regime. These solutions constitute
an extension of existing non-equilibrium supersonic Marshak wave solutions
which are linear, to the nonlinear regime, which prevails in realistic
high energy density systems. The generalized solutions assume a material
model with power law temperature dependent opacities and a material
energy density which is proportional to the radiation energy density,
as well as a surface radiation temperature drive which obeys a temporal
power-law. The solutions are analyzed in detail and it is shown that
they take various qualitatively different forms according to the values
of the opacity exponents. The solutions are used to construct a set
of standardized benchmarks for supersonic non-equilibrium radiative
heat transfer, which are nontrivial but straightforward to implement.
These solutions are compared in detail to implicit Monte-Carlo and
discrete-ordinate transport simulations as well gray diffusion simulations,
showing a good agreement, which demonstrates the usefulness of these
solutions as a code verification test problem. 
\end{abstract}
\maketitle

\section{Introduction}

Radiation hydrodynamics is paramount in the description and analysis
of high energy density systems, ranging from inertial confinement
fusion to astrophysical phenomena \cite{lindl2004physics,back2000diffusive,bailey2015higher,falize2011similarity,hurricane2014fuel,cohen2020key,heizler2021radiation}.
Analytic and semi-analytic solutions for the equations of radiation
hydrodynamics have a central role in the analysis and design of high
energy density experiments \cite{sigel1988x,lindl1995development,cohen2020key,heizler2021radiation,malka2022supersonic}
and in the process of verification and validation of computer simulations
\cite{calder2002validating,krumholz2007equations,gittings2008rage,coggeshall1991analytic,lowrie2007radiative,mcclarren2011benchmarks,bennett2021self,mcclarren2021two,mcclarren2011solutions,mcclarren2008analytic,kamm2008enhanced,rider2016robust,krief2021analytic,giron2021solutions,giron2023solutions,krief2023piston}.

The Marshak wave, developed in the seminal work \cite{marshak1958effect},
which was further generalized in Refs. \cite{petschek1960penetration,pakula1985self,kaiser1989x,reinicke1991point,shestakov1999time,hammer2003consistent,saillard2010principles,lane2013new,shussman2015full,heizler2016self,cohen2018modeling},
is a fundamental concept that arises when intense energy deposition
occurs in a material, leading to a steep temperature gradient. In
such circumstances, radiative energy transport plays a pivotal role,
and its description is far from straightforward. The Marshak wave
problem poses complex questions related to the sudden energy deposition,
the subsequent thermalization of the material, and the rapid emission
and transport of radiative energy. At high enough temperatures, the
radiative heat wave propagates faster than the speed of sound, which
results in a supersonic Marshak wave \cite{hammer2003consistent,shussman2015full,malka2022supersonic},
for which the material motion can be neglected. The original Marshak
wave problem assumes local thermodynamic equilibrium (LTE) between
the radiation field and the heated material, which is usually only
valid at long times or for optically thick systems. Pomraning \cite{pomraning1979non}
and subsequently Su and Olson \cite{bingjing1996benchmark} developed
a widely used solution for the non-equilibrium Marshak wave problem,
by defining a material model for which the heat transfer is linear.
To that end, they assumed a temperature independent opacity, a material
energy density which is proportional to the radiation energy density
and a time independent radiation temperature drive. Bennett and McClarren
have recently developed a non-linear benchmark whose results are obtained
numerically from detailed simulations \cite{bennett2023benchmark}.
There are currently no known exact solutions to the non-equilibrium
Marshak wave problem in the nonlinear regime, which prevails in most
high-energy-density systems.

In this work we develop new solutions to the non-equilibrium supersonic
Marshak wave problem, which are nonlinear and self-similar. The radiative
transfer model used is the gray, non-equilibrium diffusion approximation.
We assume a material model with power law temperature dependent opacities,
which is a good approximation for many real world materials in a wide
range of thermodynamic conditions. We also assume a material energy
density which is proportional to the radiation energy density. It
is shown that nonlinear self-similar solutions exist for a surface
radiation temperature drive which obeys a temporal power-law, which
is related to the absorption opacity exponent. We use the generalized
solutions to define a family of standardized benchmarks for supersonic
non-equilibrium radiative heat transfer. These benchmarks are compared
in detail to numerical transport and gray diffusion computer simulations.

\section{Statement of the problem\label{sec:Statement-of-the}}

In supersonic high energy density flows, where radiation heat conduction
dominates and hydrodynamic motion is negligible, the material density
is constant in time, and the heat flow is supersonic. Under these
conditions, neglecting hydrodynamic motion, the non-equilibrium 1-group
(gray) radiative transfer problem in planar slab symmetry for the
radiation energy density $E\left(x,t\right)$ and the material energy
density $u\left(x,t\right)$, is formulated by the following coupled
equations \cite{zeldovich1967physics,pomraning1982radiation,pomraning2005equations,mihalas1999foundations,castor2004radiation}:

\begin{equation}
\frac{\partial E}{\partial t}+\frac{\partial F}{\partial x}=ck_{a}\left(U-E\right),\label{eq:main_eq}
\end{equation}
\begin{equation}
\frac{\partial u}{\partial t}=ck_{a}\left(E-U\right),\label{eq:main_mat}
\end{equation}
where $c$ is the speed of light, $k_{a}$ the radiation absorption
macroscopic cross section (which is also referred to as the absorption
coefficient or absorption opacity), and $U=aT^{4}$ where $T$ is
the material temperature and $a=\frac{8\pi^{5}k_{B}^{4}}{15h^{3}c^{3}}=\frac{4\sigma}{c}$
is the radiation constant. The effective radiation temperature $T_{r}$
is related to the radiation energy density by $E=aT_{r}^{4}$. In
the diffusion approximation of radiative transfer, which is applicable
for optically thick media, the radiation energy flux obeys Fick's
law: 
\begin{equation}
F=-D\frac{\partial E}{\partial x},\label{eq:fick}
\end{equation}
where the radiation diffusion coefficient is given by:

\begin{equation}
D=\frac{c}{3k_{t}},\label{eq:diffusion_coeff}
\end{equation}
where $k_{t}=\rho\kappa_{R}$ is the total (absorption+scattering)
macroscopic transport cross section, which we also refer to as the
total opacity coefficient, where $\kappa_{R}$ the Rosseland mean
opacity and $\rho$ is the (time independent) material mass density.

The initial conditions are of a cold material and no radiation field:
\begin{equation}
E\left(x,t=0\right)=U\left(x,t=0\right)=0.\label{eq:init_cond}
\end{equation}
As for the boundary conditions, we consider an imposed surface radiation
temperature at $x=0$ which obeys a temporal power law: 
\begin{equation}
T_{r}\left(x=0,t\right)\equiv T_{s}\left(t\right)=T_{0}t^{\tau},\label{eq:Tbound}
\end{equation}
so that the radiation energy density at the system left boundary is:

\begin{equation}
E\left(x=0,t\right)=E_{0}t^{4\tau},\ E_{0}=aT_{0}^{4}.\label{eq:bc}
\end{equation}
We note that this boundary condition of an imposed surface temperature
is different than the common Marshak (or Milne) boundary condition
which is employed in the non-equilibrium Marshak wave problem \cite{pomraning1979non,bingjing1996benchmark},
which represents the flux incoming from a heat bath with a prescribed
temperature. Nevertheless, as will be shown in section \ref{subsec:Marshak-boundary-condition},
a closed form relation exists between the surface radiation temperature
and the heat bath temperature, so that the problem can be defined
alternatively by a Marshak boundary condition with a prescribed bath
temperature as a function of time. This approach was previously employed
for LTE waves in Refs. \cite{cohen2020key,heizler2021radiation,cohen2018modeling,rosen2005fundamentals}.

In this work, we assume a power law temperature dependence of the
total opacity, which we write as:

\begin{equation}
k_{t}\left(T\right)=k_{0}T^{-\alpha},\label{eq:ross_opac_powerlaw}
\end{equation}
and define a similar form for the absorption opacity: 
\begin{equation}
k_{a}\left(T\right)=k'_{0}T^{-\alpha'}.\label{eq:planck_opac_powerlaw}
\end{equation}
We note that the form \ref{eq:ross_opac_powerlaw} is equivalent to
the common power law representation \cite{hammer2003consistent,garnier2006self,smith2010solutions,shussman2015full,heizler2016self,hristov2018heat,heizler2021radiation,krief2021analytic,malka2022supersonic}
of the Rosseland opacity $\kappa_{R}\left(T,\rho\right)=\frac{1}{\mathcal{G}}T^{-\alpha}\rho^{\lambda}$,
with the coefficient $\mathcal{G}=\rho^{\lambda+1}/k_{0}$. Similarly,
the absorption opacity \ref{eq:planck_opac_powerlaw} is equivalent
to the form $k_{a}\left(T,\rho\right)=\frac{1}{\mathcal{G}'}T^{-\alpha'}\rho^{\lambda'+1}$
with $\mathcal{G}'=\rho^{\lambda'+1}/k'_{0}$.

It will be noted in Appendix \ref{sec:Dimensional-analysis} that
in order to obtain self-similar solutions to the problem at hand,
we need to assume that $u$ is proportional to $U$, that is, there
should be a quartic temperature dependence for the material energy
density. We write this in the common form: 
\begin{equation}
u\left(T\right)=\frac{aT^{4}}{\epsilon},\label{eq:eos}
\end{equation}
where $\epsilon$ is a dimensionless constant which represents the
ratio between the radiation and material energies, at equilibrium.
The form \ref{eq:eos} is a special case of the general power law
$u\left(T,\rho\right)=\mathcal{F}T^{\beta}\rho^{1-\mu}$ (see Refs.
\cite{hammer2003consistent,garnier2006self,smith2010solutions,shussman2015full,heizler2016self,hristov2018heat,heizler2021radiation,krief2021analytic,malka2022supersonic}),
with $\mathcal{F}=\frac{a}{\epsilon}$, $\beta=4$ and $\mu=1$. The
temperature dependence in equation \ref{eq:eos} is the same as employed
in the theory of linear Marshak waves \cite{pomraning1979non,bingjing1996benchmark}.

Using the material model defined in equations \ref{eq:ross_opac_powerlaw}-\ref{eq:eos},
equations \ref{eq:main_eq}-\ref{eq:main_mat} are written in closed
form as a set of nonlinear coupled partial differential equations
for $E$ and $U$:

\begin{equation}
\frac{\partial E}{\partial t}=K\frac{\partial}{\partial x}\left(U^{\frac{\alpha}{4}}\frac{\partial E}{\partial x}\right)+MU^{-\frac{\alpha'}{4}}\left(U-E\right),\label{eq:Eform}
\end{equation}

\begin{equation}
\frac{\partial U}{\partial t}=\epsilon MU^{-\frac{\alpha'}{4}}\left(E-U\right),\label{eq:Uform}
\end{equation}
where we have defined the dimensional constants: 
\begin{equation}
K=\frac{c}{3a^{\frac{\alpha}{4}}k_{0}},\label{eq:KDEF}
\end{equation}
\begin{equation}
M=ca^{\frac{\alpha'}{4}}k'_{0}.\label{eq:MDEF}
\end{equation}

\section{A Self-Similar solution\label{sec:Self-Similar-solution}}

As shown in detail in Appendix \ref{sec:Dimensional-analysis}, the
problem defined by the nonlinear gray diffusion model in equations
\ref{eq:Eform}-\ref{eq:Uform} with the initial and boundary conditions
\ref{eq:init_cond},\ref{eq:bc}, has a self-similar solution, only
for the specific surface temperature exponent give by: 
\begin{equation}
\tau=\frac{1}{\alpha'},\label{eq:tau_ss}
\end{equation}
for which the following quantity is a dimensionless constant: 
\begin{equation}
\mathcal{A}=ME_{0}^{-\frac{\alpha'}{4}}=ck'_{0}T_{0}^{-\alpha'}.\label{eq:adef}
\end{equation}
Under the constraint \ref{eq:tau_ss}, the problem can be solved using
the method of dimensional analysis \cite{buckingham1914physically,zeldovich1967physics,barenblatt1996scaling,shussman2015full,krief2021analytic},
resulting in a self-similar solution which is expressed in terms of
self-similar profiles: 
\begin{equation}
E\left(x,t\right)=E_{0}t^{4\tau}f\left(\xi\right),\label{eq:Ess}
\end{equation}
\begin{equation}
U\left(x,t\right)=E_{0}t^{4\tau}g\left(\xi\right),\label{eq:Uss}
\end{equation}
with the dimensionless similarity coordinate: 
\begin{equation}
\xi=\frac{x}{t^{\delta}\left(KE_{0}^{\frac{\alpha}{4}}\right)^{\frac{1}{2}}},\label{xsi_def}
\end{equation}
where the similarity exponent is: 
\begin{equation}
\delta=\frac{1}{2}\left(1+\tau\alpha\right)=\frac{1}{2}\left(1+\frac{\alpha}{\alpha'}\right).\label{eq:delta_def}
\end{equation}
The radiation and material temperature profiles are: 
\begin{equation}
T_{r}\left(x,t\right)=T_{0}t^{\tau}f^{1/4}\left(\xi\right),\label{eq:Trss}
\end{equation}
\begin{equation}
T\left(x,t\right)=T_{0}t^{\tau}g^{1/4}\left(\xi\right).\label{eq:Tmss}
\end{equation}
By plugging the self-similar form \ref{eq:Ess}-\ref{eq:Uss} into
the nonlinear gray diffusion system \ref{eq:Eform}-\ref{eq:Uform}
and using the relations $\frac{\partial\xi}{\partial t}=-\frac{\delta\xi}{t}$
and $\frac{\partial\xi}{\partial x}=\frac{\xi}{x}$, all dimensional
quantities are factored out, and the following (dimensionless) second
order ordinary differential equations (ODE) system for the similarity
profiles is obtained: 
\begin{align}
 & 4\tau f\left(\xi\right)-\delta f'\left(\xi\right)\xi\nonumber \\
 & =g^{\frac{\alpha}{4}-1}\left(\xi\right)\left(g\left(\xi\right)f''\left(\xi\right)+\frac{\alpha}{4}f'\left(\xi\right)g'\left(\xi\right)\right)\nonumber \\
 & -\mathcal{A}g^{-\frac{\alpha'}{4}}\left(\xi\right)\left(f\left(\xi\right)-g\left(\xi\right)\right),\label{eq:f_ode}
\end{align}
\begin{equation}
4\tau g\left(\xi\right)-\delta g'\left(\xi\right)\xi=\epsilon\mathcal{A}g^{-\frac{\alpha'}{4}}\left(\xi\right)\left(f\left(\xi\right)-g\left(\xi\right)\right).\label{eq:g_ode}
\end{equation}
The surface radiation temperature boundary condition (equation \ref{eq:bc}),
is written in terms of the radiation energy similarity profile as:
\begin{equation}
f\left(0\right)=1.\label{eq:f0_bc}
\end{equation}
It is evident that the dimensionless problem defined by equations
\ref{eq:f_ode}-\ref{eq:f0_bc} depends only on the dimensionless
parameters $\alpha$, $\alpha'$, $\epsilon$ and $\mathcal{A}$.

We assume that the solution has a steep nonlinear heat front, that
is, $f\left(\xi\right)=g\left(\xi\right)=0$ for $\xi\geq\xi_{0}$,
where $\xi_{0}$ is finite and represents the similarity coordinate
at the heat front. According to equation \ref{xsi_def}, the heat
front propagates in time according to: 
\begin{equation}
x_{F}\left(t\right)=\xi_{0}t^{\delta}\left(KE_{0}^{\frac{\alpha}{4}}\right)^{\frac{1}{2}}.\label{eq:xheat}
\end{equation}
By using equations \ref{eq:KDEF},\ref{eq:delta_def}, and \ref{eq:diffusion_coeff}
the heat front can be written as: 
\begin{align}
x_{F}\left(t\right)= & \xi_{0}\sqrt{\frac{ct}{3k_{0}\left(T_{0}t^{\tau}\right)^{-\alpha}}}=\xi_{0}\sqrt{D_{s}\left(t\right)t},
\end{align}
which is essentially a generalization of the familiar diffusion law
$x_{F}\propto\sqrt{Dt}$, with a time dependent diffusion coefficient
$D_{s}\left(t\right)=c/3k_{t}\left(T_{s}\left(t\right)\right)$ evaluated
at the (time dependent) surface radiation temperature (see also Refs.
\cite{garnier2006self,shussman2015full}). Since opacities of plasmas
usually decrease in magnitude with temperature, that is, $\alpha>0,\ \alpha'>0$,
we always have $\delta>\frac{1}{2}$ and the heat propagates faster
than classical diffusion: a result which is due to the temporal increase
of the surface temperature (as $\tau=\frac{1}{\alpha'}>0$), which
in turn increases the characteristic diffusion coefficient. According
to equation \ref{eq:delta_def}, we have an accelerating heat front
($\delta>1$) if $\alpha'<\alpha$, a constant speed front ($\delta=1$)
if $\alpha=\alpha'$ and a decelerating heat front ($\delta<1)$ if
$\alpha<\alpha'$. In the limit $\alpha\ll\alpha'$, the change in
surface temperature does not increase the characteristic diffusion
coefficient, and we recover the classical diffusion $x_{F}\propto\sqrt{t}$
propagation law, which is the familiar behavior of LTE Marshak waves
with a constant driving temperature \cite{marshak1958effect,petschek1960penetration,mihalas1999foundations,hammer2003consistent,castor2004radiation,lane2013new,heizler2016self}.

\begin{figure*}[t]
\begin{centering}
\includegraphics[scale=0.38]{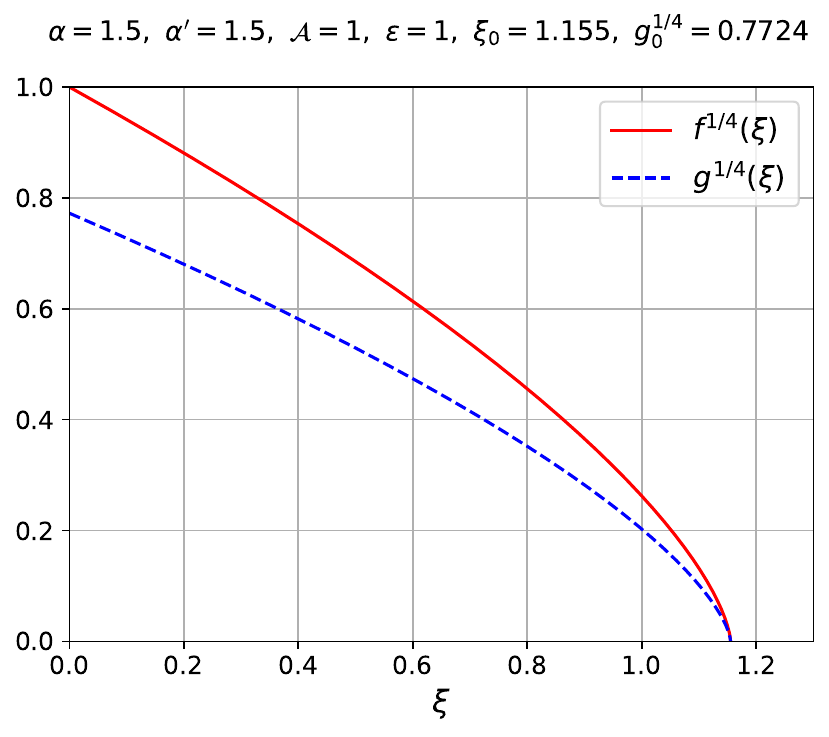}\includegraphics[scale=0.38]{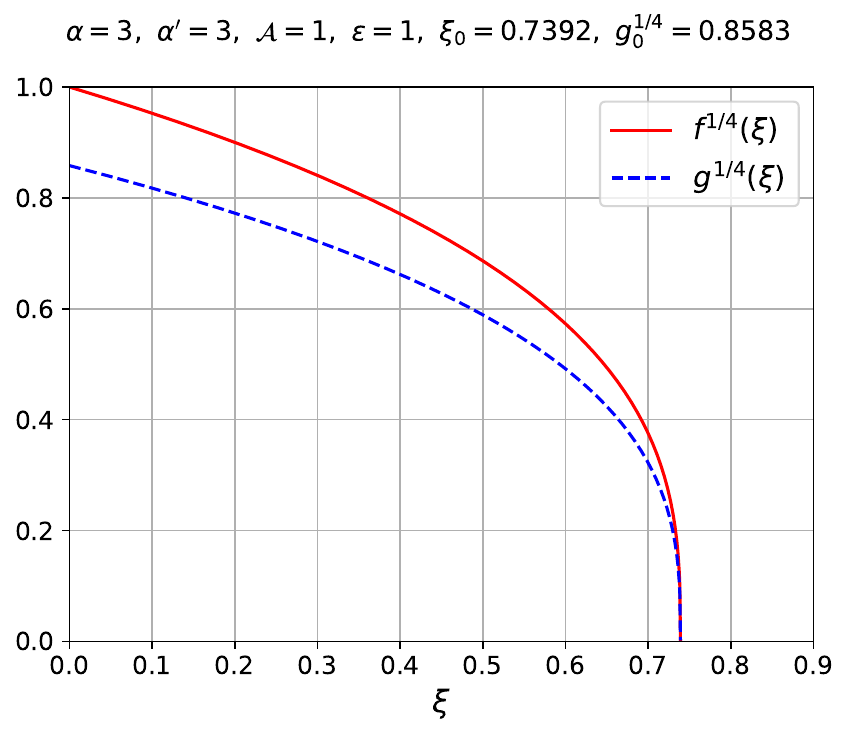}\includegraphics[scale=0.38]{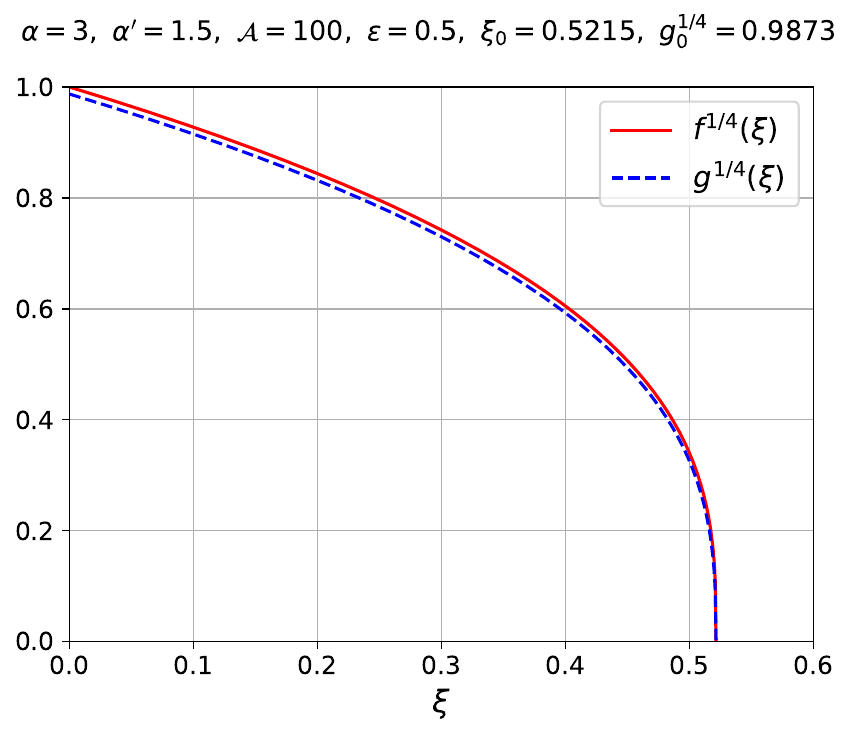} 
\par\end{centering}
\begin{centering}
\includegraphics[scale=0.38]{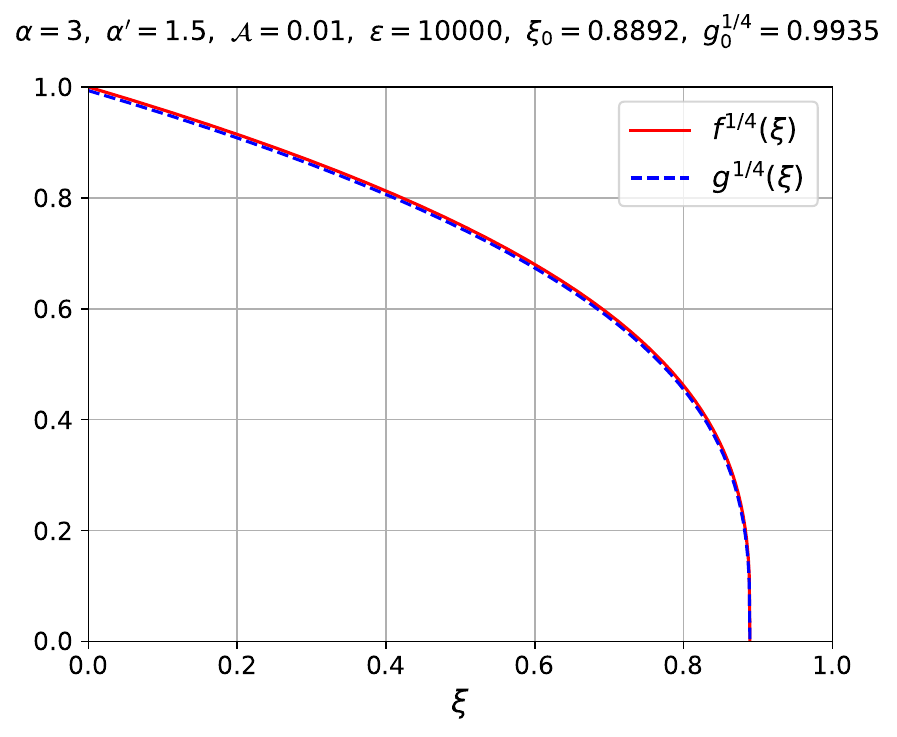}\includegraphics[scale=0.38]{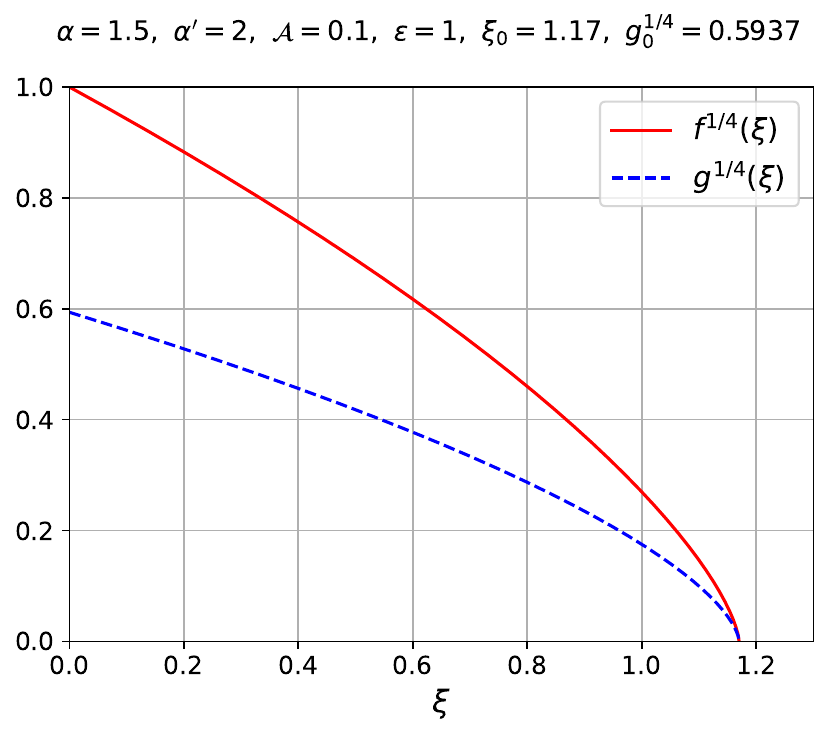}\includegraphics[scale=0.38]{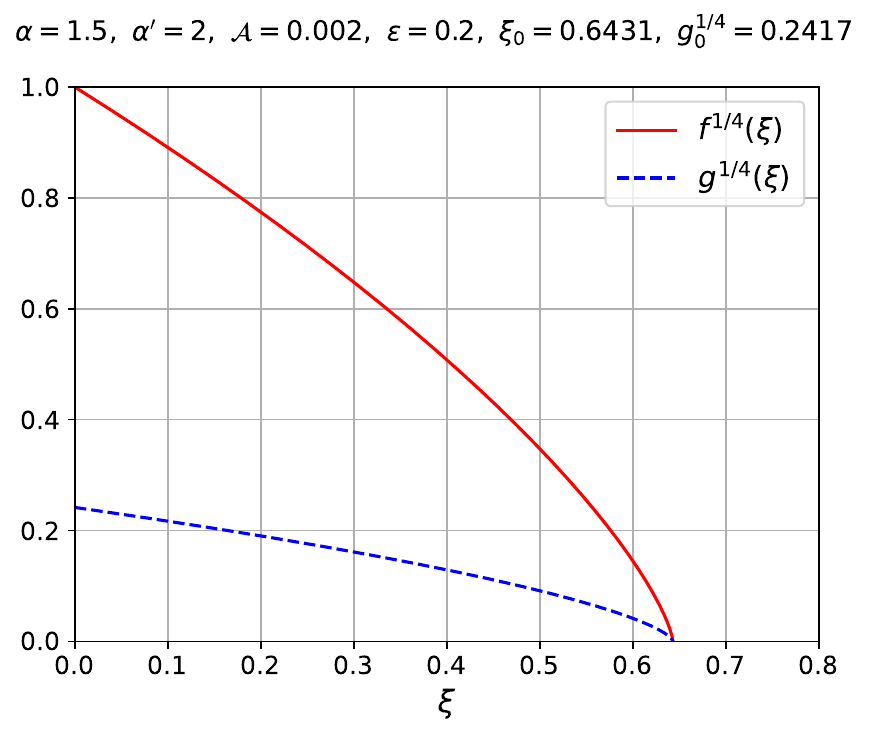} 
\par\end{centering}
\begin{centering}
\includegraphics[scale=0.38]{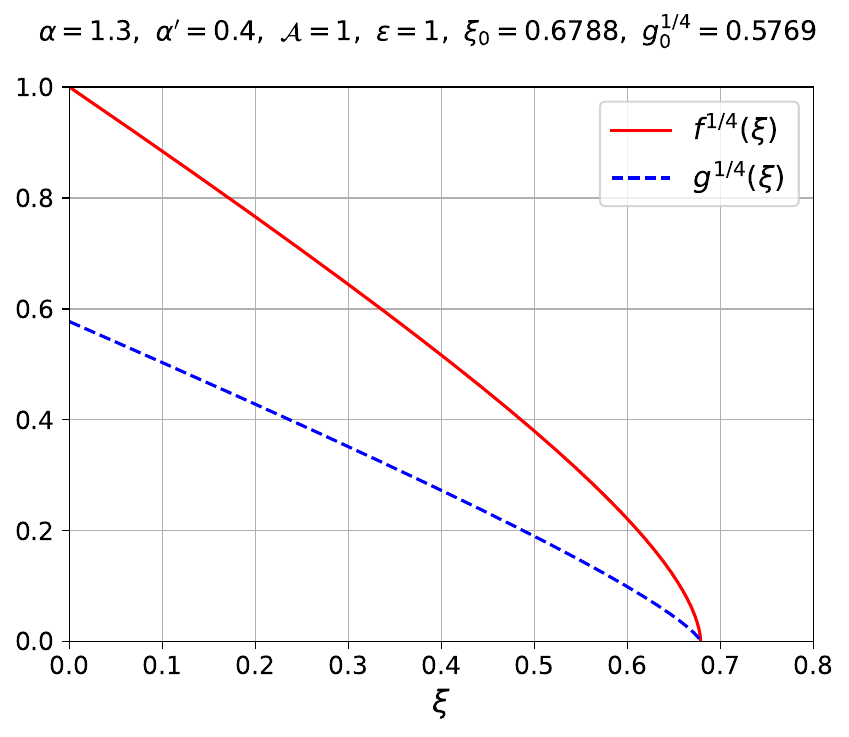}\includegraphics[scale=0.38]{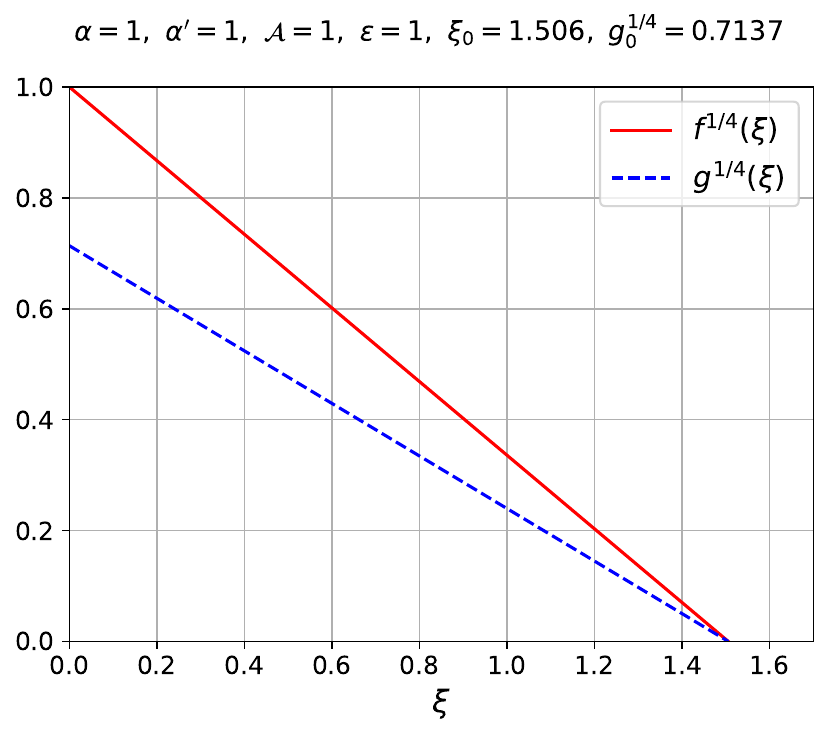}\includegraphics[scale=0.38]{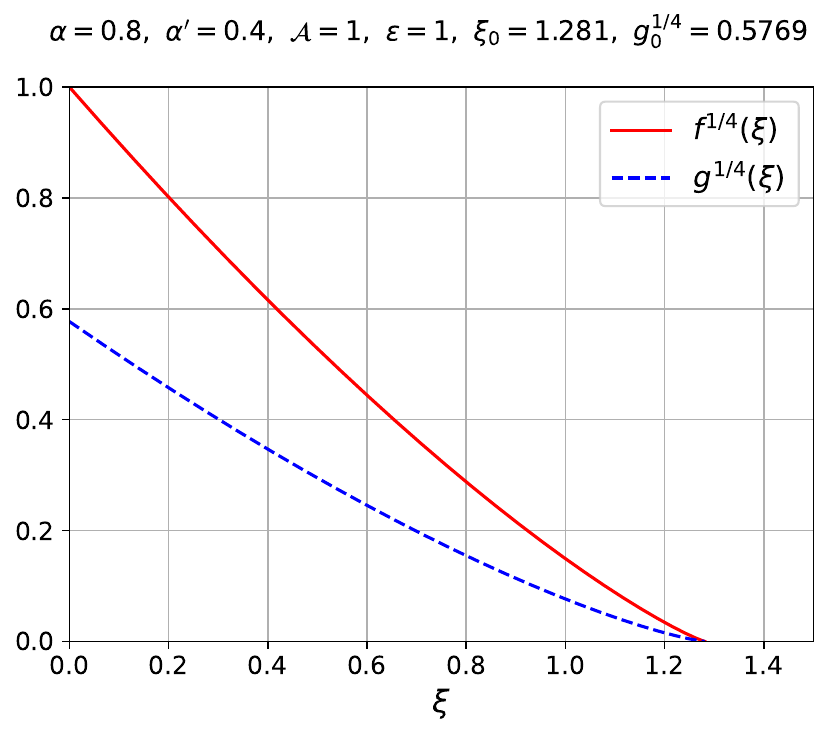} 
\par\end{centering}
\caption{Various solutions for the radiation $f^{1/4}\left(\xi\right)$ (in
red) and matter $g^{1/4}\left(\xi\right)$ (in blue) temperature similarity
profiles. The dimensionless defining parameters, namely the total
and absorption opacity temperature powers $\alpha$ and $\alpha'$,
the heat capacity ratio $\epsilon$ and the coupling parameter $\mathcal{A}$,
are listed in the figure titles together with the resulting heat front
coordinate $\xi_{0}$ and temperatures ratio at the origin, $g_{0}^{1/4}$.
As discussed in the text, it is evident that the radiation and matter
temperature become close only if $\epsilon\mathcal{A}$ is large.
\label{fig:ss_profiles}}
\end{figure*}

The value of $\xi_{0}$ is obtained by iterations of a ``shooting
method'', applied on the numerical solution of the ODE system \ref{eq:f_ode}-\ref{eq:g_ode},
which is integrated inwards from a trial $\xi_{0}$ to $\xi=0$. The
iterations adjust the trial $\xi_{0}$ until the result obeys the
boundary condition from equation \ref{eq:f0_bc} at $\xi=0$. This
is essentially the same procedure which is employed in the integration
of LTE Marshak waves \cite{marshak1958effect,petschek1960penetration,castor2004radiation,mihalas1999foundations,nelson2009semi,lane2013new,shussman2015full}.
In Fig. \ref{fig:ss_profiles}, the similarity radiation and matter
temperature profiles $f^{1/4}\left(\xi\right)$, $g^{1/4}\left(\xi\right)$
are presented for various cases.

We note that the self-similar solution \ref{eq:Ess}-\ref{eq:Uss}
is a special solution of the radiative transfer problem defined by
equations \ref{eq:main_eq}-\ref{eq:eos}, for which the ratio between
the material and radiation temperatures is time independent at any
point $\xi\propto x/x_{F}\left(t\right)$, since $T\left(x,t\right)/T_{r}\left(x,t\right)=g^{1/4}\left(\xi\right)/f^{1/4}\left(\xi\right)$
depends on $x,t$ only through the dimensionless coordinate $\xi\propto x/x_{F}\left(t\right)$.
This means that the LTE limit, for which $T\left(x,t\right)\rightarrow T_{r}\left(x,t\right)$,
is never reached, even at long times. This behavior occurs for the
unique value of the temporal exponent $\tau=1/\alpha'$ (equation
\ref{eq:tau_ss}), which enables a self-similar solution, by setting
the temporal rate in which radiation energy enters the system via
the boundary condition \ref{eq:bc} (or equivalently, as will be discussed
below, equation \ref{eq:Tbath_marsh_bc}), such that the ratio between
the radiation and matter energies are constant in time. We note that
numerical gray diffusion simulations of the radiative transfer problem
\ref{eq:main_eq}-\ref{eq:eos} with $\tau\neq1/\alpha'$, for which
the solutions are not self-similar, indeed show that the ratio $T\left(x,t\right)/T_{r}\left(x,t\right)$
is a function of time and space, and not a simple function of $\xi\propto x/x_{F}\left(t\right)$.
Moreover, it will be shown below in section \ref{subsec:An-exact-analytic},
that for the special case $\alpha=\alpha'$ the ratio $g^{1/4}\left(\xi\right)/f^{1/4}\left(\xi\right)$
does not depend on $\xi$ as well, so that the ratio between radiation
and matter temperatures is the same at all times and at all points
$\xi\propto x/x_{F}\left(t\right)$ across the heat wave.

\subsection{The LTE limit\label{subsec:The-LTE-limit}}

The dimensionless constant $\mathcal{A}$ quantifies the strength
of the emission-absorption process. If $\bar{t}$ is a typical timescale
in the problem, the typical absorption coefficient is $\bar{k}_{a}=k'_{0}\left(T_{0}\bar{t}^{\tau}\right)^{-\alpha'}$,
and from equation \ref{eq:adef} we find $\mathcal{A}=c\bar{k}_{a}\bar{t}$,
which is the typical absorption-emission energy exchange rate, multiplied
by the typical timescale. In general, the matter equation \ref{eq:main_mat}
can be written as:

\begin{align}
\frac{\partial U}{\partial t} & =c\epsilon\mathcal{A}\left(\frac{U}{E_{0}}\right)^{-\frac{\alpha'}{4}}\left(E-U\right),\label{eq:ueq}
\end{align}
which shows that the quantity $\epsilon\mathcal{A}$ determines the
equilibration rate. This is to be expected, since even for strong
emission-absorption ($\mathcal{A}\gg1$), a material with a large
heat capacity ($\epsilon\ll1$) can remain out of equilibrium, as
it takes a long time to heat the material. On the other hand, even
for a low emission-absorption rate ($\mathcal{A}\ll1$), a material
with a small heat capacity ($\epsilon\gg1$) can reach equilibrium,
as it takes a short time to be heated. Therefore, the LTE limit which
prevails in the case of strong radiation-matter coupling, should be
reached when $\epsilon\mathcal{A}\gg1$. The fact that the quantity
$\epsilon\mathcal{A}$ determines how close the problem is to LTE,
is also evident in figures \ref{fig:ss_profiles},\ref{fig:mesh_temp_rat_eps_A}
and \ref{fig:mesh_temp_rat}. In addition, equation \ref{eq:g_ode}
shows that when $\epsilon\mathcal{A}\gg1$ we have $f\left(\xi\right)\approx g\left(\xi\right)$,
that is, the radiation and material temperatures are approximately
equal, and local equilibrium is reached.

We now consider the case of strong emission-absorption ($\mathcal{A}\gg1$)
and finite $\epsilon$, such that $\epsilon\mathcal{A}\gg1$. The
total energy density is $E+u=E+\frac{1}{\epsilon}U=E_{0}t^{4\tau}h\left(\xi\right)$
with the total energy density similarity profile $h\left(\xi\right)=f\left(\xi\right)+\frac{1}{\epsilon}g\left(\xi\right)$,
which at LTE is simply $h\left(\xi\right)=\left(1+\frac{1}{\epsilon}\right)f\left(\xi\right)$,
and therefore obeys boundary condition $h\left(0\right)=1+\frac{1}{\epsilon}$.
By summing the ODEs \ref{eq:f_ode}-\ref{eq:g_ode} in the limit $f\left(\xi\right)\approx g\left(\xi\right)$,
a single second order ODE for $h\left(\xi\right)$ is obtained. This
equation, along with equation \ref{xsi_def} is equivalent to the
LTE Marshak wave problem (see equations 3 and 10 in Ref. \cite{garnier2006self}
or equations 6 and 9 in Ref. \cite{shussman2015full}), in the specific
case of a temperature surface temperature $T_{0}t^{\tau}$ with $\tau=\frac{1}{\alpha'}$
and a power law equation of state which includes the material and
radiation energies, that is $u\left(T,\rho\right)=\mathcal{F}T^{\beta}\rho^{1-\mu}$
with $\mathcal{F}=a\left(1+\frac{1}{\epsilon}\right)$, $\beta=4$
and $\mu=1$.

\subsection{The solution near the origin\label{subsec:The-solution-near}}

\begin{figure}
\begin{centering}
\includegraphics[scale=0.55]{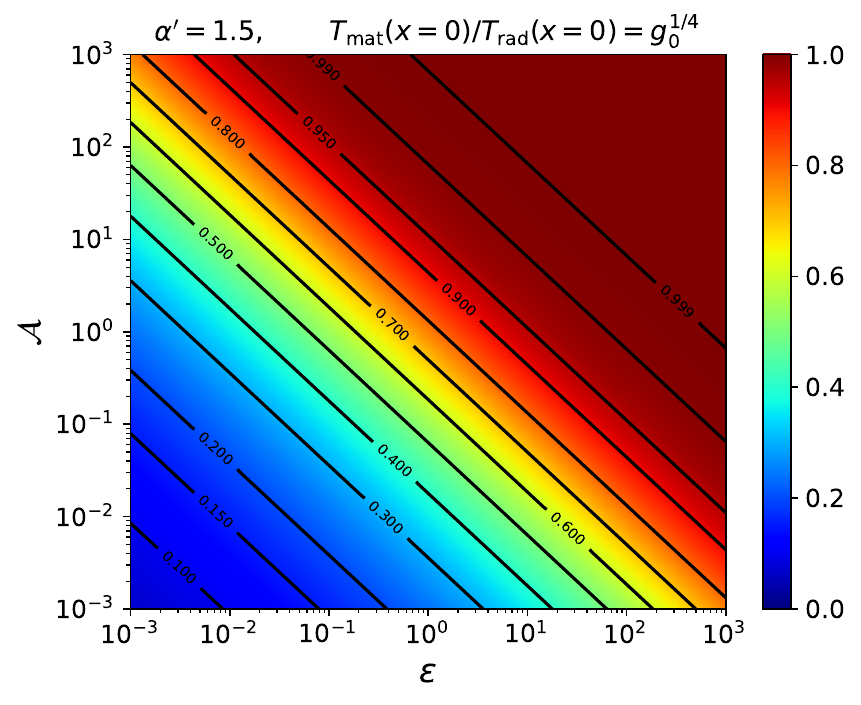} 
\par\end{centering}
\begin{centering}
\includegraphics[scale=0.55]{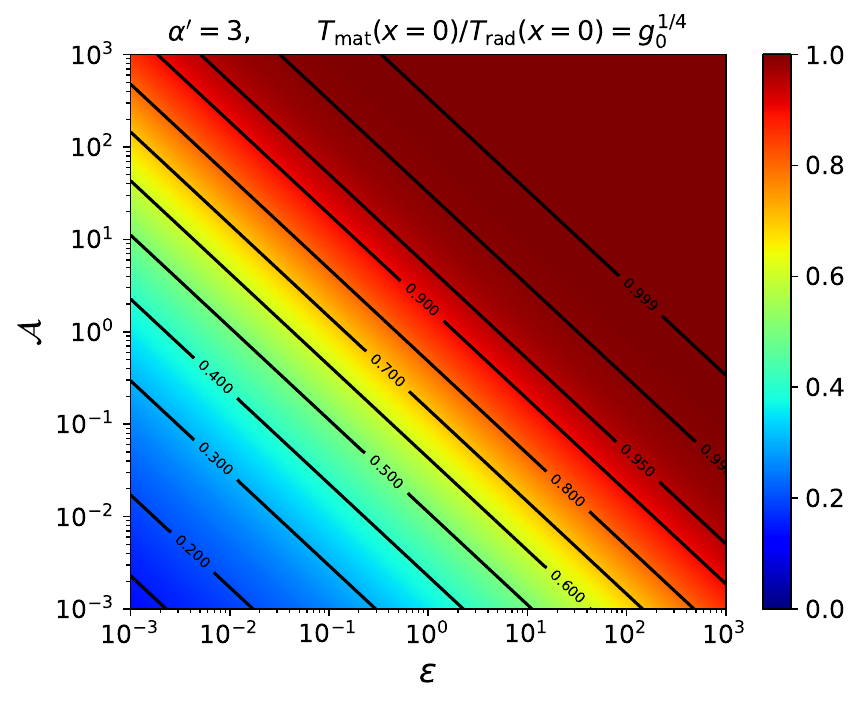} 
\par\end{centering}
\caption{The matter-radiation temperature ratio at the origin, as a function
of $\epsilon$ and $\mathcal{A}$ for the absorption opacity temperature
powers $\alpha'=1.5$ (upper pane) and $\alpha'=3$ (lower pane).
\label{fig:mesh_temp_rat_eps_A}}
\end{figure}

\begin{figure}
\begin{centering}
\includegraphics[scale=0.55]{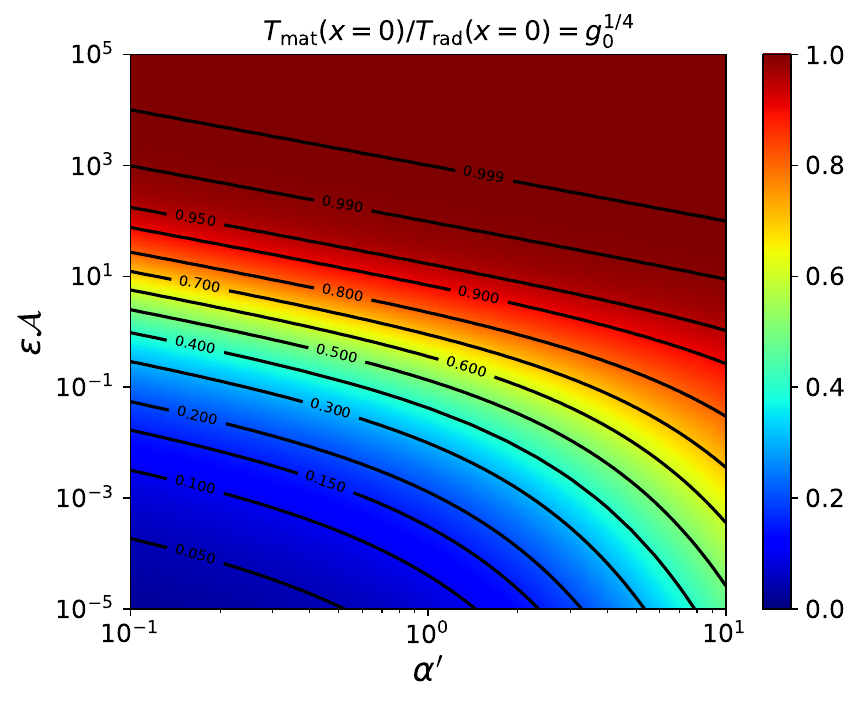} 
\par\end{centering}
\caption{The matter-radiation temperature ratio at the origin, as a function
of the absorption opacity temperature power $\alpha'$ and $\epsilon\mathcal{A}$.
\label{fig:mesh_temp_rat}}
\end{figure}

As we now show, the behavior of the solution near the system boundary,
can be analyzed without having to solve the full coupled ODE system
\ref{eq:f_ode}-\ref{eq:g_ode}. To that end, we write the first order
expansion of the solution near $\xi\rightarrow0$:

\begin{align}
f\left(\xi\right) & \approx f\left(0\right)+f'\left(0\right)\xi+O\left(\xi^{2}\right),\label{eq:expansion_origing_1}\\
g\left(\xi\right) & \approx g\left(0\right)+g'\left(0\right)\xi+O\left(\xi^{2}\right).\label{eq:expansion_origing_2}
\end{align}
Since $f\left(0\right)=1$, the value of $g_{0}\equiv g\left(0\right)$
is a measure of the deviation from equilibrium at the system's boundary:
\begin{equation}
\frac{T\left(x=0,t\right)}{T_{r}\left(x=0,t\right)}=\left(\frac{g\left(0\right)}{f\left(0\right)}\right)^{\frac{1}{4}}=g_{0}^{1/4}.
\end{equation}
By substituting the expansion \ref{eq:expansion_origing_1}-\ref{eq:expansion_origing_2}
into the matter equation \ref{eq:g_ode} and equating the zero order
terms, we find:

\begin{equation}
g_{0}\left(1+\frac{4}{\alpha'\epsilon\mathcal{A}}g_{0}^{\frac{\alpha'}{4}}\right)=1.\label{eq:g0_eq_general}
\end{equation}
Similarly, equating the first order terms gives a direct relation
between the ratio of derivatives and $g_{0}$: 
\begin{equation}
\frac{g'\left(0\right)}{f'\left(0\right)}=\frac{g_{0}}{1+\frac{1-\frac{\alpha}{\alpha'}}{2\epsilon\mathcal{A}}g_{0}^{\frac{\alpha'}{4}+1}}.\label{eq:gpfp_rat}
\end{equation}
equation \ref{eq:g0_eq_general} is a nonlinear equation for $g_{0}$
that can be solved numerically by standard root finding algorithms,
such as the Newton-Rapshon method.

First we note that in the LTE limit, since $\epsilon\mathcal{A}\gg1$,
it is evident from equation \ref{eq:g0_eq_general} that $g_{0}\approx1$
and from Eq ~\ref{eq:gpfp_rat} that $f'\left(0\right)\approx g'\left(0\right)$,
as expected, since in the LTE limit we have $f\left(\xi\right)\approx g\left(\xi\right)$.
In the general non-LTE case it is evident from equation \ref{eq:g0_eq_general}
that $g_{0}$ depends on $\alpha'$ while it does not depend on $\alpha$
at all, and that it depends on $\mathcal{A}$ and $\epsilon$ only
through their product $\epsilon\mathcal{A}$. This is to be expected,
since the matter-radiation equilibration process is dictated by the
absorption opacity rather than the total opacity (which determines
the spatial heat propagation), and since the radiation-matter coupling
rate scales as the inverse material heat capacity, which is proportional
to $\epsilon$ (as discussed in section \ref{subsec:The-LTE-limit}).
We also note that it is evident from equation \ref{eq:g0_eq_general},
that $g_{0}$ is strictly monotonic with respect to $\alpha'$, which
is to be expected since a stronger absorption temperature dependence
leads to a faster equilibration. These trends can be seen in Fig.
\ref{fig:ss_profiles}, as well as in Fig. \ref{fig:mesh_temp_rat_eps_A}
where we show the resulting solutions for $g_{0}^{1/4}$ as a function
of $\epsilon$ and $\mathcal{A}$ in a wide range for selected values
of $\alpha'$, and in Fig. \ref{fig:mesh_temp_rat} as a function
of $\alpha'$ and $\epsilon\mathcal{A}$.

\subsection{Marshak boundary condition\label{subsec:Marshak-boundary-condition}}

As opposed to the surface radiation temperature boundary condition
\ref{eq:bc}, the classical non-equilibrium Marshak wave problem \cite{pomraning1979non,bingjing1996benchmark},
as well as many other non-equilibrium benchmarks \cite{larsen1988grey,fleck1971implicit,densmore2012hybrid,steinberg2022multi,steinberg2023frequency,mclean2022multi,yee2017stable,zhang2023fully,olson2000diffusion},
are specified in terms of a given incoming radiative flux. The former
boundary condition is more natural to apply in the diffusion approximation,
while the latter is more natural to use in the solution of the radiation
transport equation, which has the angular surface flux as a boundary
condition (see below in section \ref{subsec:Transport-setup}). Nevertheless,
these two different boundary conditions are closely related.

The incoming flux boundary condition, also known as the Marshak boundary
condition \cite{pomraning1979non,bingjing1996benchmark,olson2000diffusion,rosen2005fundamentals}
at $x=0$ is:

\begin{equation}
\frac{4}{c}F_{\text{inc}}\left(t\right)=E\left(x=0,t\right)+\frac{2}{c}F\left(x=0,t\right),\label{eq:maesh_bc_def}
\end{equation}
where $F_{\text{inc}}\left(t\right)$ is a given time-dependent incoming
flux at $x=0$. For a medium coupled to a heat bath at temperature
$T_{\text{bath}}\left(t\right)$, the incoming flux is $F_{\text{inc}}\left(t\right)=\frac{ac}{4}T_{\text{bath}}^{4}\left(t\right)$.
The Marshak condition \ref{eq:maesh_bc_def} is an approximation of
the Milne boundary condition, which is valid in the diffusion limit
of radiation transport (see section \ref{subsec:Transport-setup}
below). Since the surface radiation temperature is $E\left(x=0,t\right)=aT_{s}^{4}\left(t\right)$,
the Marshak boundary condition \ref{eq:maesh_bc_def} can be written
as:

\begin{equation}
T_{\text{bath}}\left(t\right)=\left(T_{s}^{4}\left(t\right)+\frac{2}{ac}F\left(x=0,t\right)\right)^{\frac{1}{4}}.\label{eq:Tbath_marsh_bc_F}
\end{equation}
This is a statement of the Marshak boundary condition \ref{eq:maesh_bc_def},
written in terms of the bath temperature, the surface radiation temperature
and the net surface flux. For the self-similar problem considered
in this work, the radiation surface temperature is given by equation
\ref{eq:Tbound}, and by using equations \ref{eq:fick}-\ref{eq:diffusion_coeff},
the radiation flux can be written in terms of the similarity profiles:

\begin{equation}
F\left(x,t\right)=-t^{4\tau+\frac{1}{2}\left(\frac{\alpha}{\alpha'}-1\right)}K^{\frac{1}{2}}E_{0}^{1+\frac{\alpha}{8}}g^{\frac{\alpha}{4}}\left(\xi\right)f'\left(\xi\right),\label{eq:heat_flux}
\end{equation}
so that the time dependent bath temperature, according to equation
\ref{eq:Tbath_marsh_bc_F} is given by: 
\begin{equation}
T_{\text{bath}}\left(t\right)=\left(1+\mathcal{B}t^{\frac{1}{2}\left(\frac{\alpha}{\alpha'}-1\right)}\right)^{\frac{1}{4}}T_{0}t^{\tau},\label{eq:Tbath_marsh_bc}
\end{equation}
where we have defined the bath constant: 
\begin{align}
\mathcal{B} & =-\frac{2}{c}\left(KE_{0}^{\frac{\alpha}{4}}\right)^{\frac{1}{2}}g^{\frac{\alpha}{4}}\left(0\right)f'\left(0\right)\nonumber \\
 & =-\frac{2}{\sqrt{3ck_{0}T_{0}^{-\alpha}}}g^{\frac{\alpha}{4}}\left(0\right)f'\left(0\right).\label{eq:Bdef}
\end{align}
It is evident that only for $\alpha=\alpha'$ the bath temperature
is given by a temporal power law, which has the same temporal power
$\tau$ of the surface temperature. This special case will be discussed
in detail in section \ref{subsec:An-exact-analytic}. It is also evident
that $\mathcal{B}$ decreases as $k_{0}^{-1/2}$, so that $T_{\text{bath}}\left(t\right)\approx T_{s}\left(t\right)$
for optically opaque problems.

The results above agree with several previous works on LTE waves (see
for example equation 6 in Ref. \cite{cohen2020key}, equations 25-26
in Ref. \cite{heizler2021radiation} and Refs. \cite{rosen2005fundamentals,cohen2018modeling}).

\subsection{An exact analytic solution for the case $\alpha=\alpha'$ - the non-LTE
Henyey wave\label{subsec:An-exact-analytic}}

It is shown in Appendix \ref{sec:eaxactsol}, that for the special
case where the absorption coefficient and the Rosseland opacity have
the same temperature dependence, that is, when $\alpha=\alpha'$,
equations \ref{eq:f_ode}-\ref{eq:g_ode} have a simple exact analytic
solution of the form: 
\begin{equation}
f\left(\xi\right)=\left(1-\frac{\xi}{\xi_{0}}\right)^{\frac{4}{\alpha}},\label{eq:f_anal_Case}
\end{equation}

\begin{equation}
g\left(\xi\right)=g_{0}\left(1-\frac{\xi}{\xi_{0}}\right)^{\frac{4}{\alpha}},\label{eq:g_anal_Case}
\end{equation}
where $g_{0}=g\left(0\right)$ is obtained by solving the nonlinear
equation \ref{eq:g0_eq_general}, and the heat front coordinate is
given by:

\begin{equation}
\xi_{0}=\frac{2g_{0}^{\frac{\alpha}{8}}}{\sqrt{\alpha\left(1+\frac{g_{0}}{\epsilon}\right)}}.\label{eq:xsi0_exact}
\end{equation}
It is evident that the solution \ref{eq:f_anal_Case}-\ref{eq:g_anal_Case}
agrees with the general relation \ref{eq:gpfp_rat}, which for $\alpha=\alpha'$
gives $g'\left(0\right)/f'\left(0\right)=g_{0}$. We also note that
the solution \ref{eq:f_anal_Case}-\ref{eq:g_anal_Case} is special
as it results in a $\xi$ independent ratio between the material and
radiation temperatures, 
\begin{equation}
\frac{g^{1/4}\left(\xi\right)}{f^{1/4}\left(\xi\right)}=g_{0}^{1/4},
\end{equation}
which does not hold in the more general case $\alpha\neq\alpha'$,
for which this ratio depends on $\xi$. Moreover, it is evident from
equation \ref{eq:Tbath_marsh_bc} that only for $\alpha=\alpha'$
the Marshak boundary condition with a prescribed bath temperature
is given by a temporal power law, which is proportional to the surface
radiation temperature: 
\begin{equation}
T_{\text{bath}}\left(t\right)=\left(1+\mathcal{B}\right)^{\frac{1}{4}}T_{s}\left(t\right)=\left(1+\mathcal{B}\right)^{\frac{1}{4}}T_{0}t^{\tau},\label{eq:Tbath_marsh_bc-1}
\end{equation}
where the bath constant $\mathcal{B}$ defined by equation \ref{eq:Bdef},
is dimensionless for $\alpha=\alpha'$. In this case $\frac{2}{c}\left(KE_{0}^{\frac{\alpha}{4}}\right)^{\frac{1}{2}}=\sqrt{\frac{4k'_{0}}{3k_{0}\mathcal{A}}}$
and the derivative $f'\left(0\right)$ can be calculated from the
solution \ref{eq:f_anal_Case}, to give the exact expression for the
bath constant: 
\begin{align}
\mathcal{B} & =\sqrt{\frac{4k'_{0}}{3k_{0}\mathcal{A}}}\xi_{0}\left(1+\frac{g_{0}}{\epsilon}\right).\label{eq:Banal}
\end{align}
As for the LTE limit of the exact solution with $\epsilon\mathcal{A}\gg1$
and finite $\epsilon$, we have $g_{0}=1$ and: 
\begin{equation}
\xi_{0}=\frac{2}{\sqrt{\alpha\left(1+\frac{1}{\epsilon}\right)}}.\label{eq:xsi0LTE}
\end{equation}
In addition, in the LTE limit $\mathcal{B}\rightarrow0$, so that
$T_{\text{bath}}\left(t\right)\rightarrow T_{s}\left(t\right)$.

Finally, we note that it is not surprising that an exact analytical
solution exists for $\alpha=\alpha'$ for which $\tau=\frac{1}{\alpha}$.
It is known that an exact analytic solution for the LTE radiation
diffusion equation, called the Henyey Marshak wave, exists for the
specific temporal exponent $\tau_{H}=\frac{1}{4+\alpha-\beta}$ (see
Sec. II-B of Ref. \cite{hammer2003consistent}, Appendix A of Ref.
\cite{malka2022supersonic} and Refs. \cite{rosen2005fundamentals,cohen2018modeling}).
Since in this work we assumed the material energy temperature power
$\beta=4$, we have $\tau_{H}=\frac{1}{\alpha}=\tau$, so our solution
for $\alpha=\alpha'$ must reproduce the exact Henyey solution in
the LTE limit. In the more general non-LTE case (a finite $\epsilon\mathcal{A}$),
the solution \ref{eq:f_anal_Case}-\ref{eq:g_anal_Case} in fact represents
a generalization of the LTE Henyey heat wave to non-LTE gray diffusion.
This generalized solution attains some characteristics of the LTE
Henyey solution: it has a constant speed heat front $x_{F}\left(t\right)\propto t$
and a material temperature profile of the form $T\left(x,t\right)\propto T_{s}\left(t\right)\left(1-x/x_{F}\left(t\right)\right)^{\frac{1}{\alpha}}$.

\subsection{The total and absorption optical depths\label{subsec:The-optical-depth}}

The total optical depth $\mathcal{T}$, which is defined as the number
of mean free paths within the heat wave is given by \cite{rosen2005fundamentals}:
\begin{align}
\mathcal{T}\left(t\right)= & \int_{0}^{x_{F}\left(t\right)}k_{t}\left(T\left(x,t\right)\right)dx.
\end{align}
Taking the material temperature from the self-similar solution \ref{eq:Tmss},
we find: 
\begin{align}
\mathcal{T}\left(t\right)= & \left(\frac{ct}{3}k_{t}\left(T_{s}\left(t\right)\right)\right)^{\frac{1}{2}}\int_{0}^{\xi_{0}}g^{-\frac{\alpha}{4}}\left(\xi\right)d\xi\nonumber \\
= & \left(\frac{c}{3}k_{0}T_{0}^{-\alpha}t^{\left(1-\frac{\alpha}{\alpha'}\right)}\right)^{\frac{1}{2}}\int_{0}^{\xi_{0}}g^{-\frac{\alpha}{4}}\left(\xi\right)d\xi.\label{eq:OPTDEPTH_exact}
\end{align}
It is evident that for a Henyey wave with $\alpha=\alpha'$ the optical
depth does not depend on time. However, in this case the solution
is given analytically by equation \ref{eq:g_anal_Case}, for which
the integral in equation \ref{eq:OPTDEPTH_exact} diverges logarithmically
due to the steep temperature decrease near the front. This result
is in agreement with the analysis in section IV of Ref. \cite{rosen2005fundamentals},
in which LTE Henyey Marshak waves are considered, assuming a material
energy density in a general temperature power law form.

A simpler estimate for the optical depth can be obtained using the
mean-free-path at the system's boundary $k_{t}\left(T\left(x,t\right)\right)\approx k_{t}\left(T\left(x=0,t\right)\right)$.
Since the actual temperature profile is decreasing, this results in
a useful lower bound for the optical depth which can be written in
the following equivalent forms:

\begin{align}
\mathcal{T}\left(t\right)\gtrsim & \ k_{t}\left(T\left(x=0,t\right)\right)x_{F}\left(t\right)\label{eq:depth_tot}\\
= & \ \xi_{0}g_{0}^{-\frac{\alpha}{4}}\sqrt{\frac{c}{3}k_{0}T_{0}^{-\alpha}}t^{\frac{1}{2}\left(1-\frac{\alpha}{\alpha'}\right)}\nonumber \\
= & \ \xi_{0}g_{0}^{-\frac{\alpha}{8}}\sqrt{\frac{ct}{3}k_{t}\left(T\left(x=0,t\right)\right)}.\nonumber 
\end{align}
Similarly, we define the absorption optical depth, which is the number
of absorption mean free paths within the heat wave: 
\begin{align}
\mathcal{T}_{a}\left(t\right)= & \int_{0}^{x_{F}\left(t\right)}k_{a}\left(T\left(x,t\right)\right)dx,
\end{align}
which for our self-similar solution is given by: 
\begin{align}
\mathcal{T}_{a}\left(t\right) & =\frac{\mathcal{A}t^{\frac{1}{2}\left(\frac{\alpha}{\alpha'}-1\right)}}{\sqrt{3ck_{0}T_{0}^{-\alpha}}}\int_{0}^{\xi_{0}}g^{-\frac{\alpha'}{4}}\left(\xi\right)d\xi\nonumber \\
 & =\frac{x_{F}\left(t\right)}{ct}\frac{\mathcal{A}}{\xi_{0}}\int_{0}^{\xi_{0}}g^{-\frac{\alpha'}{4}}\left(\xi\right)d\xi.
\end{align}
As for the total optical depth, if $\alpha=\alpha'$ the absorption
optical depth is time independent and the dimensionless integral diverges
logarithmically. Making the same approximation by taking the absorption
mean-free-path at the system's boundary $k_{a}\left(T\left(x,t\right)\right)\approx k_{a}\left(T\left(x=0,t\right)\right)$,
we obtain the following lower bound: 
\begin{align}
\mathcal{T}_{a}\left(t\right)\gtrsim\  & k_{a}\left(T\left(x=0,t\right)\right)x_{F}\left(t\right)\label{eq:depth_abs}\\
=\  & \frac{\mathcal{A}\xi_{0}g_{0}^{-\frac{\alpha'}{4}}}{\sqrt{3ck_{0}T_{0}^{-\alpha}}}t^{\frac{1}{2}\left(\frac{\alpha}{\alpha'}-1\right)}\nonumber \\
=\  & \frac{x_{F}\left(t\right)}{ct}\mathcal{A}g_{0}^{-\frac{\alpha'}{4}}\nonumber 
\end{align}
As expected, $\mathcal{T}_{a}$ increases with $\mathcal{A}$, and
in the LTE limit, as $\mathcal{A}\rightarrow\infty$, we get $\mathcal{T}_{a}\rightarrow\infty$.
The ratio between the typical total and absorption optical depths
is given by:

\begin{align}
\frac{\mathcal{T}\left(t\right)}{\mathcal{T}_{a}\left(t\right)} & =\frac{k_{0}}{k'_{0}}\left(g_{0}^{1/4}T_{0}\right)^{\alpha'-\alpha}t^{1-\frac{\alpha}{\alpha'}}.\label{eq:depth_rat}
\end{align}
Finally, we write the \textit{effective} optical depth (see equation
1.98 in Ref. \cite{rybicki1991radiative}): 
\begin{equation}
\mathcal{T}_{\text{eff}}\equiv\sqrt{\mathcal{T}\left(t\right)\mathcal{T}_{a}\left(t\right)}\gtrsim\sqrt{\frac{\mathcal{A}}{3}}\xi_{0}\left(1+\frac{4}{\alpha'\epsilon\mathcal{A}}g_{0}^{\frac{\alpha'}{4}}\right)^{\frac{\alpha+\alpha'}{8}},\label{eq:depth_eff}
\end{equation}
which is evidently time-independent. The effective optical depth sets
the overall thermalization rate, that is, LTE is reached if and only
if $\mathcal{T}_{\text{eff}}\gg1$. This means that in general, LTE
can be reached even when $\mathcal{T}_{a}$ is not large, in a highly
scattering medium with $\mathcal{T}\gg1$ such that $\mathcal{T}_{\text{eff}}=\sqrt{\mathcal{T}\mathcal{T}_{a}}\gg1$.
This can be understood in the framework of random walk, where a large
number of scattering events increase the time between absorption events.
However, equation \ref{eq:depth_eff} shows that for the specific
problem defined in section \ref{sec:Statement-of-the}, $\mathcal{T}_{\text{eff}}$
is large only if $\mathcal{A}$ is large and therefore, LTE will be
reached only if $\mathcal{T}_{a}$ is large.

In summary, the absorption and total optical thicknesses are essentially
independent. Specifically, it is possible to define a heat wave which
is optically thick but thin with respect to absorption, that is, the
diffusion approximation holds, but the matter and radiation temperatures
are significantly different. This will be demonstrated in Sec.~\ref{sec:simulations}.

\section{Comparison with simulations\label{sec:simulations}}

\subsection{Transport setup\label{subsec:Transport-setup}}

In this section we construct a setup for a transport calculation of
the diffusion problem defined in Sec. \ref{sec:Statement-of-the}.
The general one dimensional, one group (gray) radiation transport
equation in slab symmetry for the radiation intensity field $I\left(x,\mu,t\right)$
is given by \cite{su1997analytical,olson2000diffusion,pomraning2005equations,steinberg2022multi,castor2004radiation,mihalas1999foundations,bennett2023benchmark}:

\begin{align}
\left(\frac{1}{c}\frac{\partial}{\partial t}+\mu\frac{\partial}{\partial x}\right)I\left(x,\mu,t\right)+ & \left(k_{a}\left(T\right)+k_{s}\left(T\right)\right)I\left(x,\mu,t\right)\nonumber \\
=\frac{ac}{4\pi}k_{a}\left(T\right)T^{4}\left(x,t\right)+\frac{1}{2}k_{s} & \left(T\right)\int_{-1}^{1}d\mu'I\left(x,\mu',t\right),\label{eq:Treq}
\end{align}
where $\mu$ is the directional angle cosine, $k_{a}\left(T\right)$
and $k_{s}\left(T\right)$, are, respectively, the absorption and
elastic scattering macroscopic cross sections. This transport equation
for the radiation field is coupled to the material via the following
material energy equation:

\begin{align}
\frac{\partial u\left(T\right)}{\partial t} & =k_{a}\left(T\right)\left[2\pi\int_{-1}^{1}d\mu'I\left(x,\mu',t\right)-acT^{4}\left(x,t\right)\right].\label{eq:tr_mat}
\end{align}
The radiation energy density is given by the zeroth angular moment
of the intensity via: 
\begin{equation}
E\left(x,t\right)=\frac{2\pi}{c}\int_{-1}^{1}d\mu'I\left(x,\mu',t\right),
\end{equation}
and the effective radiation temperature is $T_{r}\left(x,t\right)=\left(E\left(x,t\right)/a\right)^{1/4}$.
For optically thick problems (when the optical depth $\mathcal{T\gg}1$,
see Sec. \ref{subsec:The-optical-depth}), the diffusion limit holds,
and the transport problem \ref{eq:Treq}-\ref{eq:tr_mat} can be approximated
by the gray diffusion problem defined by equations \ref{eq:main_eq}-\ref{eq:fick},
with the total opacity $k_{t}\left(T\right)=k_{s}\left(T\right)+k_{a}\left(T\right)$.
Hence, a transport setup of the diffusion problem defined in Sec.
\ref{sec:Statement-of-the} should have the following effective elastic
scattering opacity: 
\begin{align}
k_{s}\left(T\right) & =k_{t}\left(T\right)-k_{a}\left(T\right)\nonumber \\
 & =k_{0}T^{-\alpha}-k_{0}'T^{-\alpha'}.\label{eq:scatt}
\end{align}
We note that unless $\alpha=\alpha'$, this temperature dependence
of the scattering opacity does model well real materials, but is used
here to construct a transport problem which is equivalent to a gray
diffusion problem with power law total and absorption opacities. It
is also important to note that since the scattering opacity must be
positive, the transport problem is well defined only if $k_{t}\left(T\right)\geq k_{a}\left(T\right)$
for the relevant temperatures in the problem. This constraint does
not have to hold for the analogous diffusion problem, which is well
defined for any $k_{t}$, $k_{a}$. We note that since opacity spectra
of mid or high-Z hot dense materials can be extremely detailed, the
total (one group ``Rosseland'') opacity which is dominated by spectral
dips near photon energies close to $3.8k_{B}T$, is in many cases
smaller than the absorption (one group ``Planck'') opacity, which
is dominated by spectral peaks near photon energy close to $2.8k_{B}T$
\cite{krief2018new,mihalas1999foundations,castor2004radiation,pomraning2005equations,krief2018star}.
In those realistic cases, an equivalent transport problem cannot be
defined.

Finally, the boundary condition for the transport problem is defined
naturally by an incident radiation field for incoming directions $\mu>0$,
which is given by a black body radiation bath: 
\begin{equation}
I\left(x=0,\mu,t\right)=\frac{ac}{4\pi}T_{\text{bath}}^{4}\left(t\right),\label{eq:Ibath_tr}
\end{equation}
where the time dependent bath temperature is given by solution of
the Diffusion problem via equation \ref{eq:Tbath_marsh_bc}, which
is obtained from the Marshak (Milne) boundary condition, as detailed
in Sec. \ref{subsec:Marshak-boundary-condition}. We note that the
Marshak boundary condition (equation \ref{eq:maesh_bc_def}) is obtained
by an angular integration of the exact boundary condition given by
equation \ref{eq:Ibath_tr}, and by assuming a first order spherical
harmonics expansion of the angular flux at the boundary, which is
equivalent to the diffusion approximation (see also equations 36-38
in Ref. \cite{olson2000diffusion} and Sec. II in Ref. \cite{rosen2005fundamentals}).

We note that since the diffusion limit is reached in optically thick
problems, when $\mathcal{T\gg}1$, it is expected that in this case
transport simulations will agree with diffusion simulations (and the
self-similar solutions). Independently, we expect the radiation and
matter to be out of equilibrium for absorption thin problems where
$\mathcal{T}_{a}\lesssim1$. We conclude that it is possible to construct
optically thick problems with dominant scattering ($k_{a}\ll k_{t}$),
which are absorption thin, for which we expect transport results to
agree with diffusion, while the radiation and matter are significantly
out of equilibrium. This reasoning is used in the construction of
the test cases below.

\subsection{Test cases}

We define six benchmarks based on the self-similar solutions and specify
in detail the setups for gray diffusion and transport computer simulations.
We have performed gray diffusion simulations as well as deterministic
discrete-ordinates ($S_{N}$) and stochastic implicit Monte-Carlo
(IMC) \cite{fleck1971implicit,mcclarren2009modified} transport simulations.
The diffusion simulations shown in this subsection were all performed
without the application of flux limiters (see also subsection \ref{subsec:Flux-limited-diffusion}).
The $S_{N}$ simulations were performed using the numerical method
detailed in \cite{mcclarren2022data}, while the IMC simulations were
performed using the novel numerical scheme which was recently developed
by Steinberg and Heizler in Refs. \cite{steinberg2022new,steinberg2022multi,steinberg2023frequency}.

The results are compared in figures \ref{fig:simulation_1}-\ref{fig:simulation_6},
where the temperature profiles are plotted at the final time as well
as at the times when the heat front reaches $20\%$ and $60\%$ of
the final front position.

The typical scales are as follows: temperatures are in keV, time in
nanoseconds and distance in centimeters. The material energy density
is given by: 
\begin{equation}
u\left(T\right)=\frac{a}{\epsilon}T^{4}=\frac{1.372017\times10^{14}}{\epsilon}\left(\frac{T}{\text{keV}}\right)^{4}\ \text{\ensuremath{\frac{\text{erg}}{\text{cm}^{3}}}}
\end{equation}
The heat front position (equation \ref{eq:xheat}) reads:

\begin{equation}
x_{F}\left(t\right)=\frac{3.1612\xi_{0}}{\sqrt{k_{0}T_{0}^{-\alpha}}}\left(\frac{t}{\text{ns}}\right)^{\frac{1}{2}\left(1+\frac{\alpha}{\alpha'}\right)}\ \text{cm},\label{eq:xheat_cm}
\end{equation}
and dimensionless coupling constant (equation \ref{eq:adef}) is:

\begin{equation}
\mathcal{A}=29.979k_{0}'T_{0}^{-\alpha'},\label{eq:Adef}
\end{equation}
where $T_{0}$ is measured in units of $\text{keV}/\text{ns}^{\frac{1}{\alpha'}}$,
$k_{0}$ in $\text{keV}^{\alpha}/\text{cm}$ and $k_{0}'$ in $\text{keV}^{\alpha'}/\text{cm}$.
All tests are run until the final time $t_{\text{end}}=1\text{ns}$,
and will have a surface radiation temperature of 1keV at the final
time, that is, we take $T_{0}=1\text{keV}/\text{ns}^{1/\alpha'}$
so that the surface radiation temperature is: 
\begin{equation}
T_{s}\left(t\right)=\left(\frac{t}{\text{ns}}\right)^{\frac{1}{\alpha'}}\ \text{keV}.\label{eq:Ts}
\end{equation}

\subsubsection{TEST 1}

\begin{figure}[t]
\begin{centering}
\includegraphics[scale=0.55]{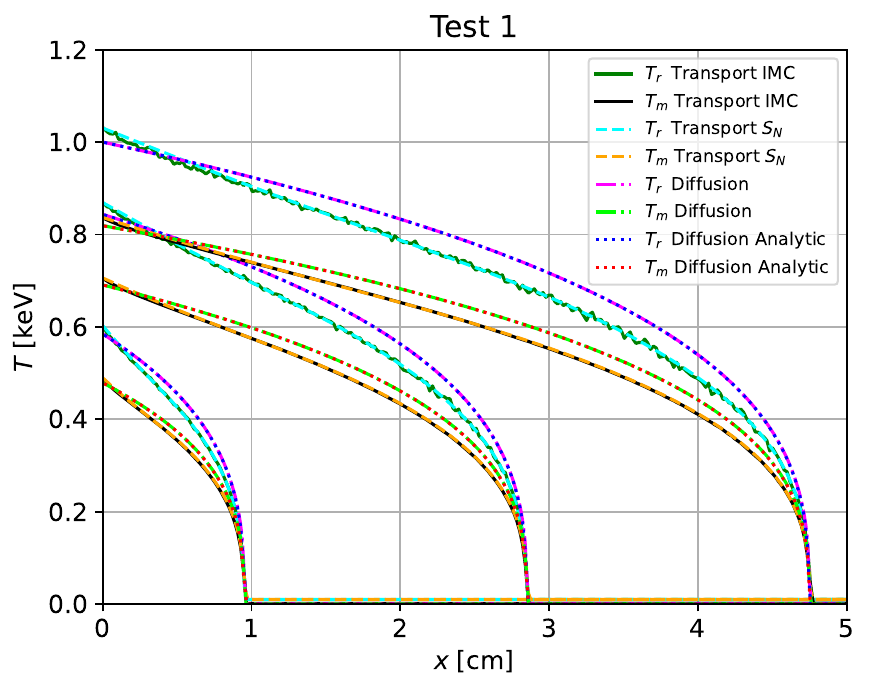} 
\par\end{centering}
\caption{Radiation and material temperature profiles for Test 1. Results are
shown at times $t=0.2,\ 0.6$ and 1ns, as obtained from a gray diffusion
simulation and from Implicit-Monte-Carlo (IMC) and discrete ordinates
($S_{N}$) transport simulations, and are compared to the analytic
solution of the gray diffusion equation (given in equations \ref{eq:xh_1},\ref{eq:Tr_anal_I}-\ref{eq:Tm_anal_I}).\label{fig:simulation_1}}
\end{figure}

In this case we take a heat capacity ratio of $\epsilon=0.2$ and
opacities with $\alpha=\alpha'=3$ and $k_{0}=k_{0}'=0.1\text{keV}^{3}/\text{cm}$,
so that the total and absorption opacities are: 
\begin{equation}
k_{t}\left(T\right)=k_{a}\left(T\right)=0.1\left(\frac{T}{\text{keV}}\right)^{-3}\ \text{cm}^{-1},
\end{equation}
which models absorption without scattering ($k_{s}\left(T\right)=0$
for transport simulations). The surface radiation temperature is:
\begin{equation}
T_{s}\left(t\right)=\left(\frac{t}{\text{ns}}\right)^{\frac{1}{3}}\ \text{keV}.
\end{equation}

\begin{figure}[t]
\begin{centering}
\includegraphics[scale=0.55]{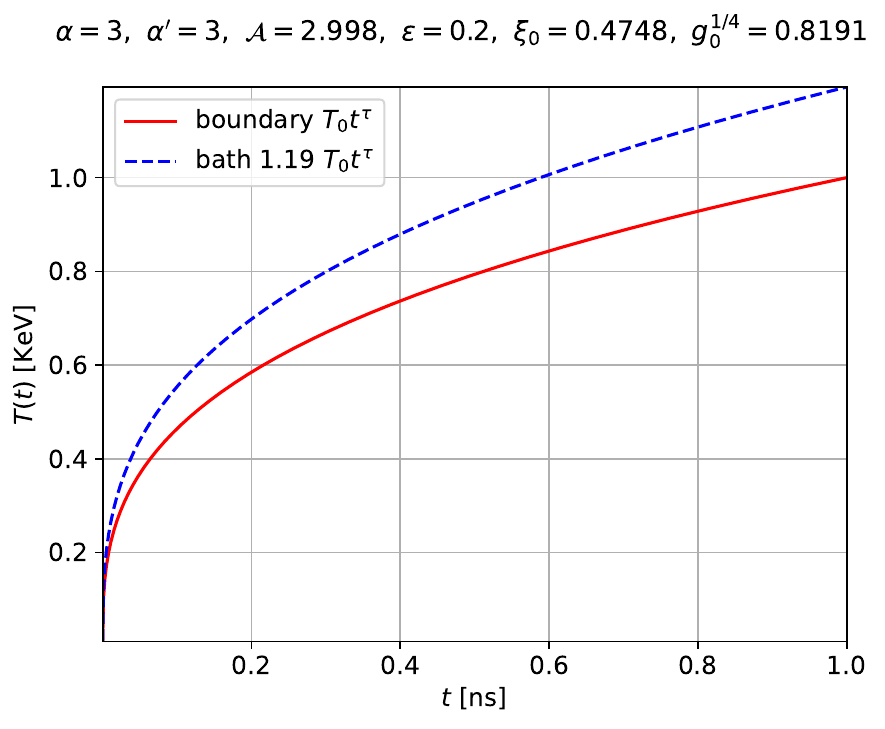} 
\par\end{centering}
\caption{A comparison between the surface (red line) and bath (dashed line)
driving temperatures for Test 1 (given in equation \ref{eq:Tabth_1}).\label{fig:simulation_1_Tbath}}
\end{figure}

For these parameters equation \ref{eq:Adef} gives the dimensionless
coupling parameter $\mathcal{A}=2.9979$. The solution of the nonlinear
equation \ref{eq:g0_eq_general} gives the boundary temperatures ratio
$g_{0}^{1/4}=0.8190643$. Since $\alpha=\alpha'$, this case is a
non-LTE Henyey wave which propagates linearly in time and has an analytical
solution as described in Sec. \ref{subsec:An-exact-analytic}. From
equation \ref{eq:xsi0_exact} we find the front coordinate $\xi_{0}=0.4747709$
which via equation \ref{eq:xheat_cm} gives the heat front position:
\begin{equation}
x_{F}\left(t\right)=4.7460665\left(\frac{t}{\text{ns}}\right)\ \text{cm}.\label{eq:xh_1}
\end{equation}
From equation \ref{eq:Banal} we find the bath constant $\mathcal{B}=1.029124$,
so that the bath temperature (equation \ref{eq:Tbath_marsh_bc-1})
is: 
\begin{equation}
T_{\text{bath}}\left(t\right)=1.193513\left(\frac{t}{\text{ns}}\right)^{\frac{1}{3}}\ \text{keV},\label{eq:Tabth_1}
\end{equation}
which is used in transport simulations via the incoming bath radiation
flux (equation \ref{eq:Ibath_tr}) or in diffusion simulations via
the Marshak boundary condition (equation \ref{eq:maesh_bc_def}).
We note that diffusion simulations can be run equivalently using the
surface temperature boundary condition (equation \ref{eq:Tbound}).
A comparison of the surface and bath temperatures as a function of
time are shown in Figure \ref{fig:simulation_1_Tbath}. The most obvious
difference between the two temperatures is that the nearly 20\% difference
between the surface temperature and the bath temperature in this case.

The radiation and matter temperature profiles are given analytically
by using equations \ref{eq:f_anal_Case}-\ref{eq:g_anal_Case} and
\ref{eq:Trss}-\ref{eq:Tmss}:

\begin{equation}
T_{r}\left(x,t\right)=\left(\frac{t}{\text{ns}}\right)^{\frac{1}{3}}\left(1-\frac{x}{x_{F}\left(t\right)}\right)^{\frac{1}{3}}\ \text{keV}\label{eq:Tr_anal_I}
\end{equation}
\begin{equation}
T\left(x,t\right)=0.8190643T_{r}\left(x,t\right).\label{eq:Tm_anal_I}
\end{equation}
Since $\alpha=\alpha'$, the total and absorption optical depths (see
equations \ref{eq:depth_tot}-\ref{eq:depth_abs}) are time independent
and given by $\mathcal{T}=\mathcal{T}_{a}\approx0.864$.

In Figure \ref{fig:simulation_1} we can see that the transport results
disagree with the analytic and numerical diffusion solutions in the
temperature profiles, but coincide with the analytic model as to the
location of the wavefront and the boundary temperatures in the figure.
This overall disagreement between transport and diffusion results
is not surprising, since this problem is not optically thick. However,
the agreement in the front position occurs since this test is opaque
enough such that diffusion theory obeys the free-streaming limit,
as discussed below in section \ref{subsec:Flux-limited-diffusion}.

\subsubsection{TEST 2}

\begin{figure}[t]
\begin{centering}
\includegraphics[scale=0.55]{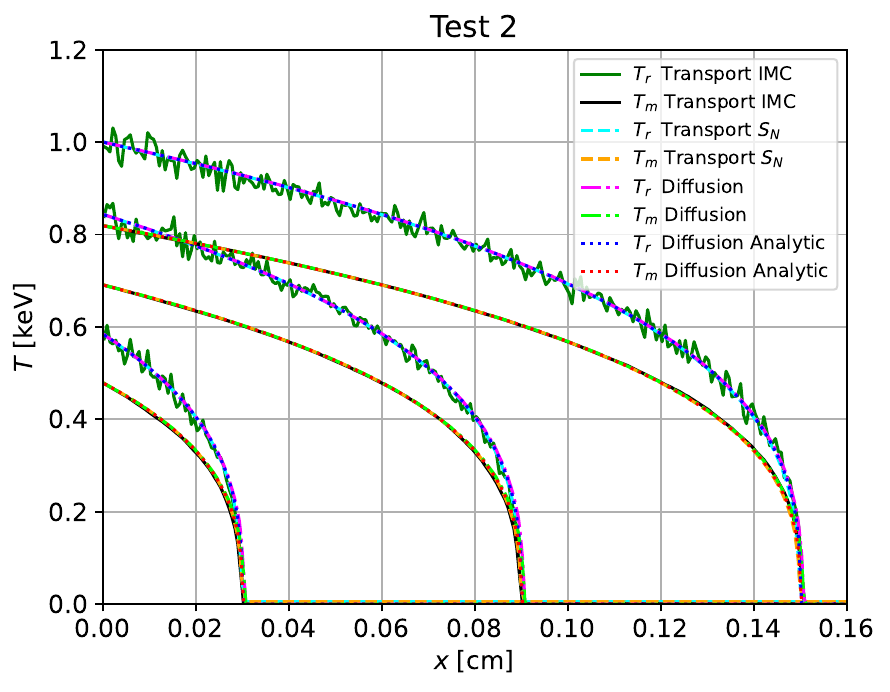} 
\par\end{centering}
\caption{Radiation and material temperature profiles for Test 2. Results are
shown at times $t=0.2,\ 0.6$ and 1ns, as obtained from a gray diffusion
simulation and from Implicit-Monte-Carlo (IMC) and discrete ordinates
($S_{N}$) transport simulations, and are compared to the analytic
solution of the gray diffusion equation (given in equations \ref{eq:Tr_anal_2}-\ref{eq:Tm_anal_2},\ref{eq:xh_2}).\label{fig:simulation_2}}
\end{figure}

This case is constructed to give the optically thick limit of Test
1, by significantly increasing the total opacity while keeping the
absorption opacity constant. As a result, we expect that the transport
calculations will agree with diffusion and therefore with the analytic
solutions. A physical mechanism for such increase is the introduction
of photon scattering, which alters the total opacity only. We take
the same parameters $\epsilon=0.2$, $\alpha=\alpha'=3$, $k_{0}'=0.1\text{\,keV}^{3}/\text{cm}$
as in Test 1, with a total opacity which is increased by a factor
of $10^{3}$ so that $k_{0}=100\,\text{keV}^{3}/\text{cm}$. Hence,
the total and absorption opacities are: 
\[
k_{t}\left(T\right)=100\left(\frac{T}{\,\text{keV}}\right)^{-3}\ \text{cm}^{-1},
\]
\[
k_{a}\left(T\right)=0.1\left(\frac{T}{\,\text{keV}}\right)^{-3}\ \text{cm}^{-1}.
\]
For transport simulations, the scattering opacity (equation \ref{eq:scatt})
is now nonzero and given by the following temperature power law: 
\[
k_{s}\left(T\right)=99.9\left(\frac{T}{\text{keV}}\right)^{-3}\ \text{cm}^{-1}.
\]
The surface radiation temperature is the same as in Test 1: 
\begin{equation}
T_{s}\left(t\right)=\left(\frac{t}{\text{ns}}\right)^{\frac{1}{3}}\ \text{keV}.
\end{equation}
Since $k_{0}'$ is the same as in Test 1, we have also the same dimensionless
coupling parameter $\mathcal{A}$. Therefore, since all dimensionless
parameters ($\alpha$, $\alpha'$, $\epsilon$ and $\mathcal{A}$)
are the same as in Test 1, this test defines the same dimensionless
problem as Test 1 with the same $g_{0}^{1/4}$, $\xi_{0}$ and Henyey
self similar profiles:

\begin{equation}
T_{r}\left(x,t\right)=\left(\frac{t}{\text{ns}}\right)^{\frac{1}{3}}\left(1-\frac{x}{x_{F}\left(t\right)}\right)^{\frac{1}{3}}\ \text{keV}\label{eq:Tr_anal_2}
\end{equation}
\begin{equation}
T\left(x,t\right)=0.8190643T_{r}\left(x,t\right).\label{eq:Tm_anal_2}
\end{equation}

However, the total optical depth, which increases with the total opacity
coefficient as $k_{0}^{1/2}$ is now larger by a factor of $\sqrt{10^{3}}\approx31.6$,
so that $\mathcal{T}\approx27.3$, while the absorption optical depth
is decreased by the same factor so that $\mathcal{T}_{a}\approx0.0273$.
Therefore, this case indeed defines an opaque heat wave so that exact
transport results should coincide with the diffusion approximation.
On the other hand, as in Test 1, the wave has a small absorption optical
depth, which results in the same significant deviation from equilibrium
as given by $g_{0}^{1/4}$. The heat front position which also depends
$k_{0}^{-1/2}$ (see equation \ref{eq:xheat_cm}), runs $\approx31.6$
times slower: 
\begin{equation}
x_{F}\left(t\right)=0.1500838\left(\frac{t}{\text{ns}}\right)\ \text{cm}.\label{eq:xh_2}
\end{equation}
Similarly, equation \ref{eq:Banal} gives a smaller bath constant
by the same factor, $\mathcal{B}=0.032544$, so that, as discussed
in Sec. \ref{subsec:Marshak-boundary-condition}, the bath temperature
is much closer to the surface temperature: 
\begin{equation}
T_{\text{bath}}\left(t\right)=1.008038\left(\frac{t}{\text{ns}}\right)^{\frac{1}{3}}\ \text{keV}.
\end{equation}

In Figure \ref{fig:simulation_2} we plot the analytic solution compared
with several numerical calculations. We observe that, modulo the stochastic
noise in the IMC calculations, all of the calculations agree within
the figure scale as expected.

\subsubsection{TEST 3}

\begin{figure}[t]
\begin{centering}
\includegraphics[scale=0.55]{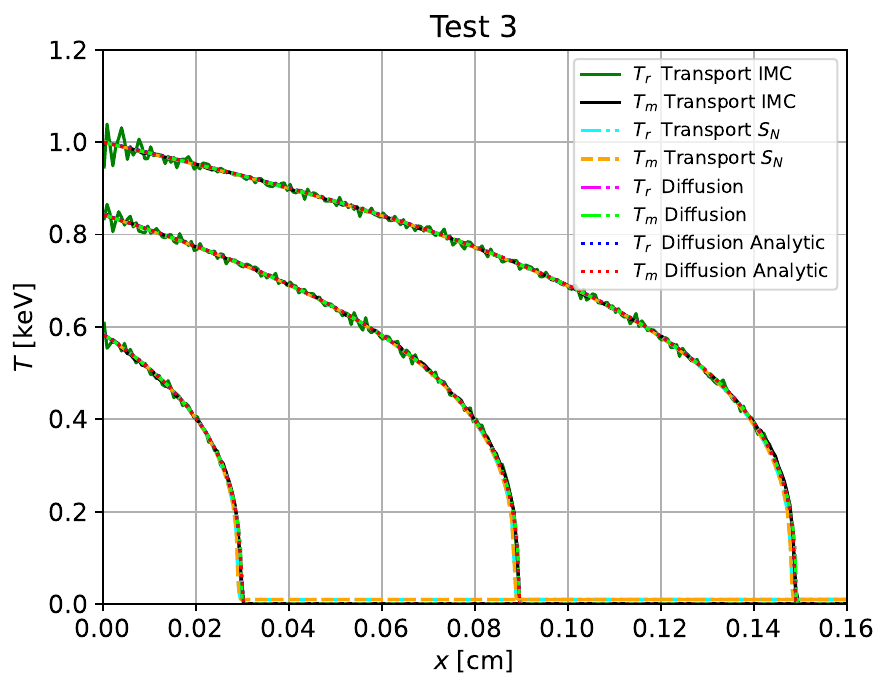} 
\par\end{centering}
\caption{Radiation and material temperature profiles for Test 3. Results are
shown at times $t=0.2,\ 0.6$ and 1ns, as obtained from a gray diffusion
simulation and from Implicit-Monte-Carlo (IMC) and discrete ordinates
($S_{N}$) transport simulations, and are compared to the analytic
solution of the gray diffusion equation (given in equations \ref{eq:xh_3},
\ref{eq:Tr_anal_3}-\ref{eq:Tm_anal_3}).\label{fig:simulation_3}}
\end{figure}

\begin{figure}[t]
\begin{centering}
\includegraphics[scale=0.47]{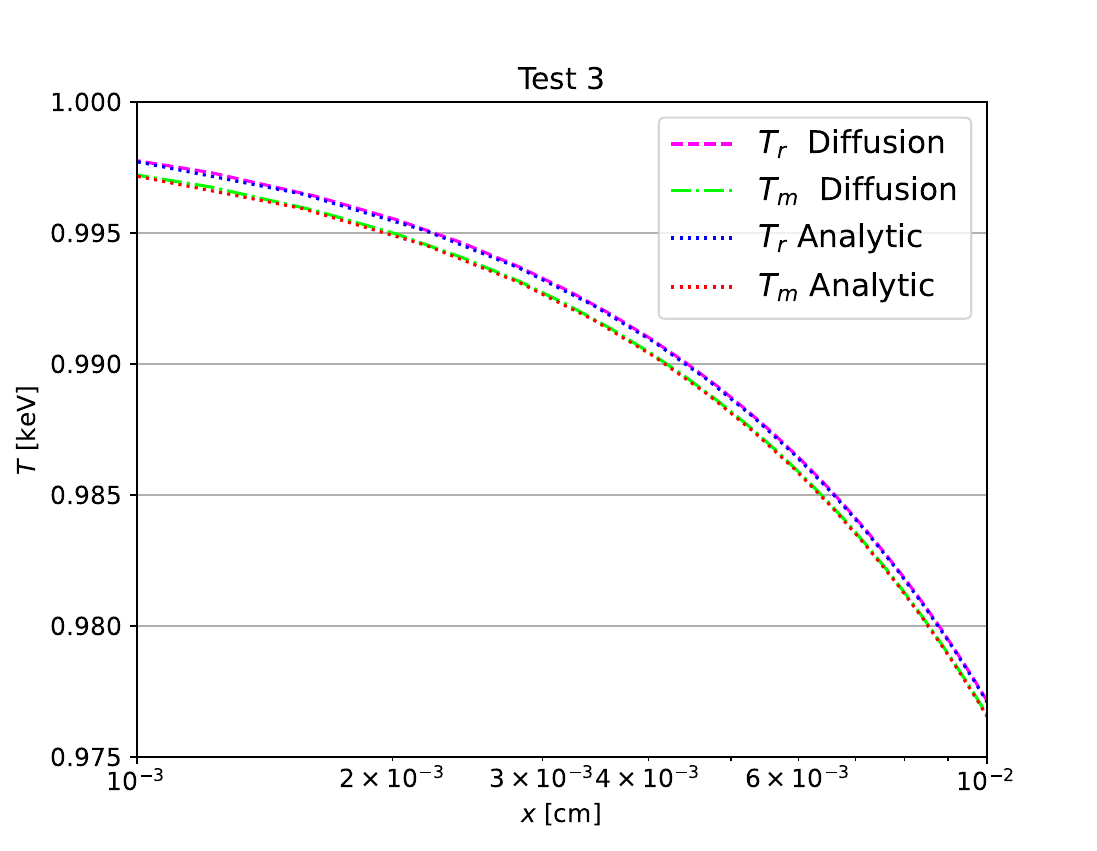} 
\par\end{centering}
\caption{A close view near origin of the temperature profiles (from diffusion
simulations and analytic solution), for Test 3 (see Fig. \ref{fig:simulation_3}).\label{fig:simulation_3_zoom}}
\end{figure}

This case is constructed to give the LTE limit of Test 2, by significantly
increasing the absorption opacity, while keeping the total opacity
constant. It is defined with the same parameters $\epsilon=0.2$,
$\alpha=\alpha'=3$, and with the absorption opacity increased by
a factor of $10^{3}$ so that $k_{0}=k_{0}'=100\text{keV}^{3}/\text{cm}$:
\[
k_{t}\left(T\right)=k_{a}\left(T\right)=100\left(\frac{T}{\text{keV}}\right)^{-3}\ \text{cm}^{-1},
\]
which as in Test 1, models pure absorption without scattering ($k_{s}\left(T\right)=0$
for transport simulations). The surface radiation temperature is the
same as in Tests 1 and 2: 
\begin{equation}
T_{s}\left(t\right)=\left(\frac{t}{\text{ns}}\right)^{\frac{1}{3}}\ \text{keV}.
\end{equation}
According to equation \ref{eq:Adef}, the dimensionless coupling parameter
should increase by a factor of $10^{3}$, so that $\mathcal{A}=2997.9$
and $\epsilon\mathcal{A}\approx600$ which is much larger then unity.
Therefore, as discussed in Sec. \ref{subsec:The-LTE-limit}, the LTE
limit should be reached. The solution of equation \ref{eq:g0_eq_general}
gives $g_{0}^{1/4}=0.999446$, so that the matter and radiation temperatures
are almost equal. From equation \ref{eq:xsi0_exact} we find the front
coordinate $\xi_{0}=0.471448$ which gives the heat front position:
\begin{equation}
x_{F}\left(t\right)=0.14903337\left(\frac{t}{\text{ns}}\right)\ \text{cm}.\label{eq:xh_3}
\end{equation}
From equation \ref{eq:Banal} we find the bath constant $\mathcal{B}=0.05954$,
so that the bath temperature is: 
\begin{equation}
T_{\text{bath}}\left(t\right)=1.014565\left(\frac{t}{\text{ns}}\right)^{\frac{1}{3}}\ \text{keV}.
\end{equation}
The radiation and matter temperature profiles are again given analytically
by:

\begin{equation}
T_{r}\left(x,t\right)=\left(\frac{t}{\text{ns}}\right)^{\frac{1}{3}}\left(1-\frac{x}{x_{F}\left(t\right)}\right)^{\frac{1}{3}}\ \text{keV}\label{eq:Tr_anal_3}
\end{equation}
\begin{equation}
T\left(x,t\right)=0.999446T_{r}\left(x,t\right).\label{eq:Tm_anal_3}
\end{equation}
Finally, the total and absorption optical depths are $\mathcal{T}=\mathcal{T}_{a}\approx14.93$.
Hence, the heat wave is optically and absorption thick, which means
that the diffusion approximation holds and that the problem is close
to LTE. We note that this optical depth is lower than Test 2 by a
factor of $0.819^{3}\approx0.55$, since the matter temperature is
now almost equal to the radiation temperature, and it was increased
by a factor of $1/0.819$ relative to Test 2.

As in Test 2, all of the numerical solutions agree with the analytic
diffusion solution, as shown in Figure \ref{fig:simulation_3}. Upon
zooming into the detail of the solutions at a small value of $x$
(see Figure \ref{fig:simulation_3_zoom}), we observe that there is
a small difference between the radiation and material temperatures
predicted by the analytic solution. This difference is captured in
the diffusion numerical solution as well.


\subsubsection{TEST 4}

\begin{figure}[t]
\begin{centering}
\includegraphics[scale=0.55]{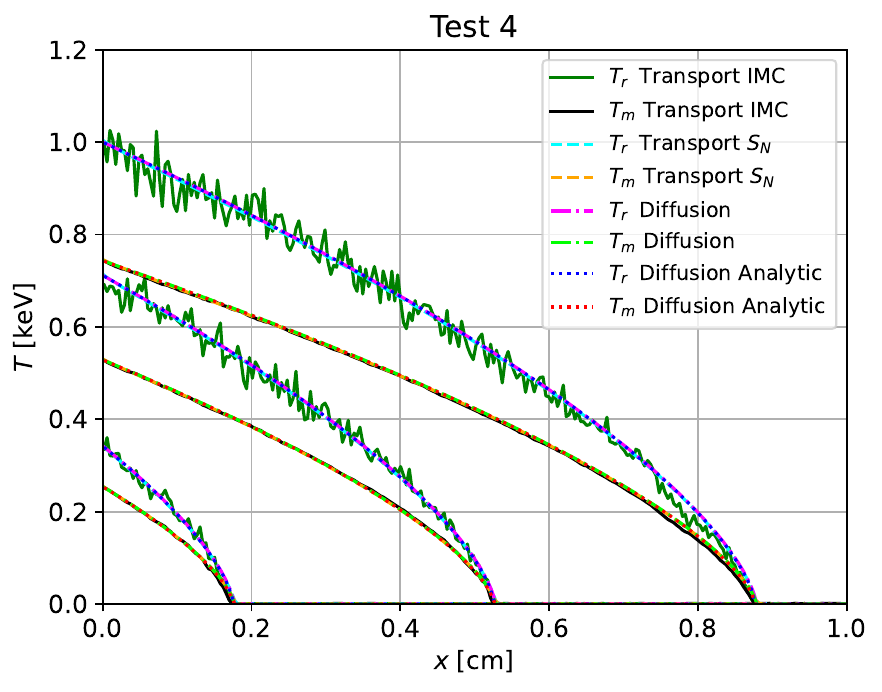} 
\par\end{centering}
\caption{Radiation and material temperature profiles for Test 4. Results are
shown at times $t=0.2,\ 0.6$ and 1ns, as obtained from a gray diffusion
simulation and from Implicit-Monte-Carlo (IMC) and discrete ordinates
($S_{N}$) transport simulations, and are compared to the analytic
solution of the gray diffusion equation (given in equations \ref{eq:xh_4},
\ref{eq:Tr_anal_4}-\ref{eq:Tm_anal_4}).\label{fig:simulation_4}}
\end{figure}

We define another non-LTE Henyey wave similar to Test 2, that is,
optically thick and absorption thin, but with a weaker opacity temperature
dependence. The resulting solution profile will approach zero more
gradually at the wave front than in previous cases. We take $\epsilon=0.25$
and smaller opacity powers $\alpha=\alpha'=1.5$ and coefficients
$k_{0}=10\text{\,keV}^{1.5}/\text{cm}$, $k_{0}'=0.1\text{\,keV}^{1.5}/\text{cm}$,
so that the total and absorption opacities are, respectively: 
\[
k_{t}\left(T\right)=10\left(\frac{T}{\text{keV}}\right)^{-3/2}\ \text{cm}^{-1},
\]
\[
k_{a}\left(T\right)=0.1\left(\frac{T}{\text{keV}}\right)^{-3/2}\ \text{cm}^{-1}.
\]
For transport simulations, the scattering opacity (equation \ref{eq:scatt})
is: 
\[
k_{s}\left(T\right)=9.9\left(\frac{T}{\text{keV}}\right)^{-3/2}\ \text{cm}^{-1}.
\]
The surface radiation temperature is: 
\begin{equation}
T_{s}\left(t\right)=\left(\frac{t}{\text{ns}}\right)^{\frac{2}{3}}\ \text{keV}.
\end{equation}
As $k_{0}'$ is the same as in tests 1 and 2, we have the same dimensionless
coupling parameter $\mathcal{A}=2.9979$. The resulting value of the
boundary temperatures ratio is $g_{0}^{1/4}=0.7431154$ and the front
coordinate is $\xi_{0}=0.877244$, so that the heat front position
reads: 
\begin{equation}
x_{F}\left(t\right)=0.876941\left(\frac{t}{\text{ns}}\right)\ \text{cm}.\label{eq:xh_4}
\end{equation}
The resulting bath constant is $\mathcal{B}=0.129865$, so that the
bath temperature is: 
\begin{equation}
T_{\text{bath}}\left(t\right)=1.030995\left(\frac{t}{\text{ns}}\right)^{\frac{2}{3}}\ \text{keV},
\end{equation}
The radiation and matter temperature profiles are given analytically
by:

\begin{equation}
T_{r}\left(x,t\right)=\left(\frac{t}{\text{ns}}\right)^{\frac{2}{3}}\left(1-\frac{x}{x_{F}\left(t\right)}\right)^{\frac{2}{3}}\ \text{keV}\label{eq:Tr_anal_4}
\end{equation}
\begin{equation}
T\left(x,t\right)=0.7431154T_{r}\left(x,t\right).\label{eq:Tm_anal_4}
\end{equation}
The problem is indeed optically thick with a total optical depth of
$\mathcal{T}\approx13.69$, while it is absorption thin with an absorption
optical depth small by a factor of $k_{0}/k_{0}'=100$ (see equation
\ref{eq:depth_rat}) so that $\mathcal{T}_{a}\approx0.1369$.

From Figure \ref{fig:simulation_4} we observe the more gradual approach
to zero of the solution near the wavefront. We also observe that in
this problem there is increased noise in the IMC solution. This is
likely due to the relatively small amount of scattering and absorption/emission
behind the wavefront in this problem.


\subsubsection{TEST 5}

\begin{figure}[t]
\begin{centering}
\includegraphics[scale=0.55]{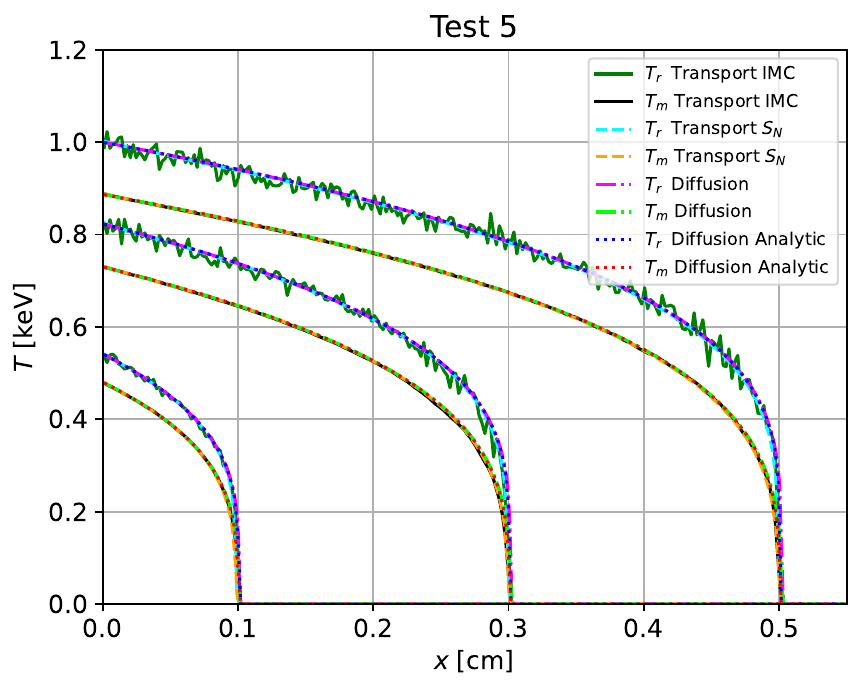} 
\par\end{centering}
\caption{Radiation and material temperature profiles for Test 5. Results are
shown at times $t=0.341995$, $0.711379$ and $1$ns, as obtained
from a gray diffusion simulation and from Implicit-Monte-Carlo (IMC)
and discrete ordinates ($S_{N}$) transport simulations, and are compared
to the semi-analytic solution of the gray diffusion equation (given
in equations \ref{eq:xheat_5}, \ref{eq:Tr_anal_5}-\ref{eq:Tm_anal_5}
and table \ref{tab:test5_6}). \label{fig:simulation_5}}
\end{figure}

\begin{table}[t]
\centering{}%
\begin{tabular}{|c|c|c|c|c|}
\cline{2-5} \cline{3-5} \cline{4-5} \cline{5-5} 
\multicolumn{1}{c|}{} & \multicolumn{2}{c|}{Test 5} & \multicolumn{2}{c|}{Test 6}\tabularnewline
\hline 
$\xi/\xi_{0}$  & $f^{1/4}\left(\xi\right)$  & $g^{1/4}\left(\xi\right)$  & $f^{1/4}\left(\xi\right)$  & $g^{1/4}\left(\xi\right)$\tabularnewline
\hline 
\hline 
0  & 1  & 0.88669  & 1  & 0.6163\tabularnewline
\hline 
0.05  & 0.9859  & 0.87304  & 0.98116  & 0.60675\tabularnewline
\hline 
0.1  & 0.97134  & 0.85891  & 0.96155  & 0.59685\tabularnewline
\hline 
0.15  & 0.95626  & 0.84425  & 0.94111  & 0.58655\tabularnewline
\hline 
0.2  & 0.94061  & 0.82901  & 0.91978  & 0.57582\tabularnewline
\hline 
0.25  & 0.92434  & 0.81312  & 0.89745  & 0.5646\tabularnewline
\hline 
0.3  & 0.90738  & 0.79651  & 0.87404  & 0.55284\tabularnewline
\hline 
0.35  & 0.88963  & 0.77909  & 0.84941  & 0.54047\tabularnewline
\hline 
0.4  & 0.871  & 0.76074  & 0.82344  & 0.52741\tabularnewline
\hline 
0.45  & 0.85135  & 0.74133  & 0.79595  & 0.51356\tabularnewline
\hline 
0.5  & 0.83053  & 0.72068  & 0.76672  & 0.49879\tabularnewline
\hline 
0.55  & 0.80832  & 0.69857  & 0.7355  & 0.48295\tabularnewline
\hline 
0.6  & 0.78447  & 0.67471  & 0.70193  & 0.46582\tabularnewline
\hline 
0.65  & 0.75859  & 0.6487  & 0.66559  & 0.44712\tabularnewline
\hline 
0.7  & 0.73016  & 0.61999  & 0.62583  & 0.42645\tabularnewline
\hline 
0.75  & 0.69842  & 0.58773  & 0.58177  & 0.40323\tabularnewline
\hline 
0.8  & 0.66213  & 0.55061  & 0.53202  & 0.3765\tabularnewline
\hline 
0.85  & 0.6191  & 0.5063  & 0.47422  & 0.34464\tabularnewline
\hline 
0.9  & 0.56479  & 0.44996  & 0.40368  & 0.30424\tabularnewline
\hline 
0.95  & 0.48623  & 0.36802  & 0.30787  & 0.2458\tabularnewline
\hline 
0.973  & 0.42851  & 0.30796  & 0.2433  & 0.20323\tabularnewline
\hline 
0.99  & 0.35349  & 0.23127  & 0.16833  & 0.14928\tabularnewline
\hline 
0.996  & 0.29886  & 0.17771  & 0.12111  & 0.11194\tabularnewline
\hline 
0.998  & 0.26442  & 0.14565  & 0.09493  & 0.089799\tabularnewline
\hline 
0.999  & 0.23464  & 0.1194  & 0.074677  & 0.071877\tabularnewline
\hline 
0.9999  & 0.1598  & 0.06176  & 0.034189  & 0.033862\tabularnewline
\hline 
0.99999  & 0.10977  & 0.031938  & 0.015817  & 0.015782\tabularnewline
\hline 
0.999999  & 0.075329  & 0.016436  & 0.0073397  & 0.0073363\tabularnewline
\hline 
\end{tabular}\caption{The temperature similarity profiles $f^{1/4}$ and $g^{1/4}$ resulting
from the numerical solutions of the ODE system \ref{eq:f_ode}-\ref{eq:g_ode},
as a function of $\xi/\xi_{0}$, for Test 5 ($\xi_{0}=0.5006965$)
and Test 6 ($\xi_{0}=0.7252338$). It is evident that the temperatures
ratio $g^{1/4}\left(\xi\right)/f^{1/4}\left(\xi\right)$ depends on
$\xi$ and is monotonically decreasing/increasing for Test 5 and Test
6, respectively.\label{tab:test5_6}}
\end{table}

This case defines an optically thick and absorption thin wave (as
in Test 2 and 4), but with $\alpha>\alpha'$. This means that in contrast
to all previous cases, the heat front will not propagate at a constant
speed and as it is not a Henyey solution, the self-similar profiles
must be calculated numerically. We take $\epsilon=1$, a total opacity
with $\alpha=3.5$, $k_{0}=10\text{\,keV}^{3}/\text{cm}$: 
\[
k_{t}\left(T\right)=10\left(\frac{T}{\text{keV}}\right)^{-3.5}\ \text{cm}^{-1},
\]
and an absorption opacity with $\alpha'=1.75$ and $k_{0}'=0.1\text{keV}^{3}/\text{cm}$:
\[
k_{a}\left(T\right)=0.1\left(\frac{T}{\text{keV}}\right)^{-1.75}\ \text{cm}^{-1}.
\]
For transport simulations, the scattering opacity (equation \ref{eq:scatt})
is now a difference between two power laws: 
\begin{equation}
k_{s}\left(T\right)=10\left(\frac{T}{\text{keV}}\right)^{-3.5}-0.1\left(\frac{T}{\text{keV}}\right)^{-1.75}\ \text{cm}^{-1}.\label{eq:ks_test5}
\end{equation}
The surface radiation temperature is: 
\begin{equation}
T_{s}\left(t\right)=\left(\frac{t}{\text{ns}}\right)^{\frac{4}{7}}\ \text{keV}.
\end{equation}
The scattering opacity \ref{eq:ks_test5} is positive for all temperatures
in the problem for short enough times, and specifically for $t\leq t_{\text{end}}=1\text{\,ns}$
(since $k_{s}\left(T_{s}\left(t_{\text{end}}\right)\right)>0$). Since
$k_{0}'$ is the same as in tests 1, 2 and 4, we have again $\mathcal{A}=2.9979$.
The solution of the nonlinear equation \ref{eq:g0_eq_general} gives
the exact boundary temperatures ratio $g_{0}^{1/4}=0.886692$. As
mentioned above, since $\alpha\neq\alpha'$, the similarity profiles
must be solved numerically by integrating the ODE system \ref{eq:f_ode}.
The resulting temperature similarity profiles are given in table \ref{tab:test5_6}.
It is evident that the tabulated numerical solution agrees with the
exact value of $g\left(0\right)/f\left(0\right)$ and the exact relation
\ref{eq:gpfp_rat} for $g'\left(0\right)/f'\left(0\right)$. The front
coordinate is found to be $\xi_{0}=0.5006965$ so that the non-linear
heat front position is given by: 
\begin{equation}
x_{F}\left(t\right)=0.50052323\left(\frac{t}{\text{ns}}\right)^{1.5}\ \text{cm},\label{eq:xheat_5}
\end{equation}
which, since $\alpha>\alpha'$, accelerates in time. The numerical
solution has $f'\left(0\right)=-2.217102$, and from equation \ref{eq:Bdef}
we find the bath constant $\mathcal{B}=0.0970622\text{\,ns}^{-1/2}$,
so that the bath temperature, which in this case is not a simple temporal
power law, is given by: 
\begin{equation}
T_{\text{bath}}\left(t\right)=\left(1+0.0970622\left(\frac{t}{\text{ns}}\right)^{\frac{1}{2}}\right)^{\frac{1}{4}}\left(\frac{t}{\text{ns}}\right)^{\frac{4}{7}}\ \text{keV}.
\end{equation}
Using the self-similar solution \ref{eq:Trss}-\ref{eq:Tmss}, the
radiation and material temperature profiles are given by: 
\begin{equation}
T_{r}\left(x,t\right)=\left(\frac{t}{\text{ns}}\right)^{\frac{4}{7}}f^{1/4}\left(\xi_{0}x/x_{F}\left(t\right)\right)\ \text{keV},\label{eq:Tr_anal_5}
\end{equation}
\begin{equation}
T\left(x,t\right)=\left(\frac{t}{\text{ns}}\right)^{\frac{4}{7}}g^{1/4}\left(\xi_{0}x/x_{F}\left(t\right)\right)\ \text{keV}.\label{eq:Tm_anal_5}
\end{equation}
It is evident from the tabulated solution that the ratio $T\left(x,t\right)/T_{r}\left(x,t\right)$
is a decreasing function of $x/x_{F}\left(t\right)$. This is in contrast
to the previous Henyey solutions (Tests 1-4) for which this ratio
is constant along the heat wave. Finally, we note that since $\alpha\neq\alpha'$,
the total and absorption optical depths lower bounds (see equations
\ref{eq:depth_tot}-\ref{eq:depth_abs}) are time dependent. Their
values at the end of the simulation are $\mathcal{T}\left(t_{\text{end}}\right)\approx7.63$
and $\mathcal{T}_{a}\left(t_{\text{end}}\right)\approx0.062$, so
that the wave is indeed optically thick and absorption thin. In figure
\ref{fig:simulation_5} the self-similar solution is compared to numerical
simulations. 

\subsubsection{TEST 6}

\begin{figure}[t]
\begin{centering}
\includegraphics[scale=0.55]{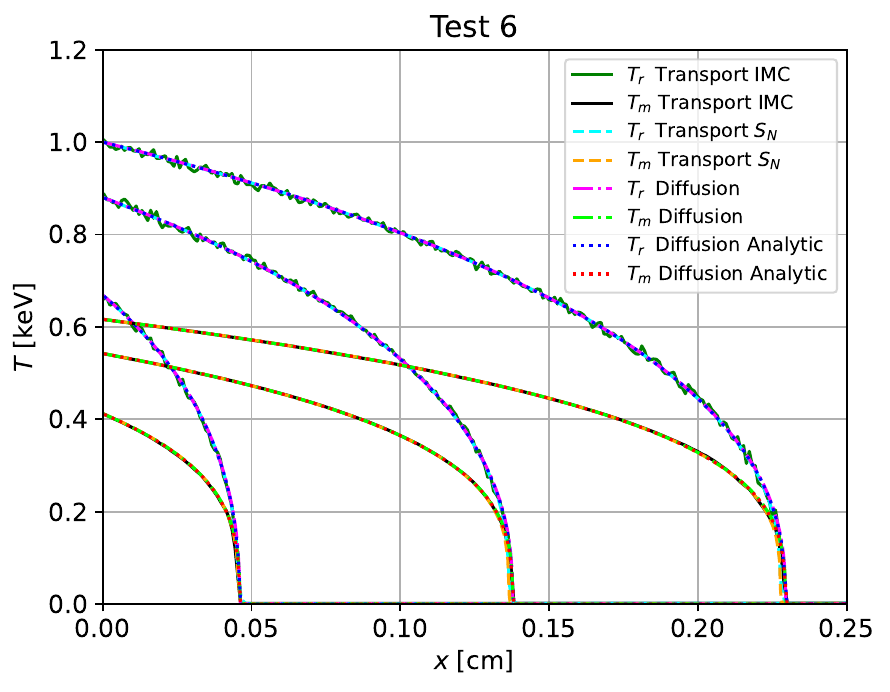} 
\par\end{centering}
\caption{Radiation and material temperature profiles for Test 6. Results are
shown at times $t=0.133748,0.528067$ and $1$ns, as obtained from
a gray diffusion simulation and from Implicit-Monte-Carlo (IMC) and
discrete ordinates ($S_{N}$) transport simulations, and are compared
to the semi-analytic solution of the gray diffusion equation (given
in equations \ref{eq:xheat_6}, \ref{eq:Tr_anal_6}-\ref{eq:Tm_anal_6}
and table \ref{tab:test5_6}).\label{fig:simulation_6}}
\end{figure}

This case defines a similar wave as in Test 5, but with $\alpha<\alpha'$.
We take $\epsilon=4$, a total opacity with $\alpha=3$ and $k_{0}=100\text{keV}^{3}/\text{cm}$:
\[
k_{t}\left(T\right)=100\left(\frac{T}{\text{keV}}\right)^{-3}\ \text{cm}^{-1},
\]
and an absorption opacity with $\alpha'=5$ and $k_{0}'=10^{-4}\text{keV}^{5}/\text{cm}$:
\[
k_{a}\left(T\right)=10^{-4}\left(\frac{T}{\text{keV}}\right)^{-5}\ \text{cm}^{-1}.
\]
For transport simulations, the scattering opacity is: 
\begin{equation}
k_{s}\left(T\right)=100\left(\frac{T}{\text{keV}}\right)^{-3}-10^{-4}\left(\frac{T}{\text{keV}}\right)^{-5}\ \text{cm}^{-1}.\label{eq:ks_test6}
\end{equation}
The surface radiation temperature is: 
\begin{equation}
T_{s}\left(t\right)=\left(\frac{t}{\text{ns}}\right)^{\frac{1}{5}}\ \text{keV}.
\end{equation}
Since $\alpha<\alpha'$, the scattering opacity is always negative
at low enough temperatures. The coefficients $k_{0},k_{0}'$ were
chosen such that the scattering opacity in equation \ref{eq:ks_test6}
will become negative only at temperatures lower than $1\text{eV}.$
Therefore, transport simulations are initialized with a ``cold''
material temperature of $1\text{eV}$, which is much lower than the
final radiation surface temperature of $1\text{keV},$and therefore,
could only cause a negligible difference between those simulations
and the analytic solution. Using equation \ref{eq:Adef}, we find
$\mathcal{A}=0.0029979$ for this case and the solution of equation
\ref{eq:g0_eq_general} gives the boundary temperatures ratio $g_{0}^{1/4}=0.6163$.
As in Test 5, the similarity temperature profiles are solved numerically
and given in table \ref{tab:test5_6}, and the front coordinate is
found to be $\xi_{0}=0.7252338$ so that the heat front position is
given by: 
\begin{equation}
x_{F}\left(t\right)=0.22925972\left(\frac{t}{\text{ns}}\right)^{0.8}\ \text{cm}.\label{eq:xheat_6}
\end{equation}
Since $\alpha<\alpha'$, the heat front decelerates in in time. The
numerical solution has $f'\left(0\right)=-2.038806$, and from equation
\ref{eq:Bdef} we find the bath constant $\mathcal{B}=0.010065\text{\,ns}^{1/5}$, and the bath temperature is given by: 
\begin{equation}
T_{\text{bath}}\left(t\right)=\left(1+0.010065\left(\frac{t}{\text{ns}}\right)^{-\frac{1}{5}}\right)^{\frac{1}{4}}\left(\frac{t}{\text{ns}}\right)^{\frac{1}{5}}\ \text{keV}.
\end{equation}
The radiation and material temperature profiles are given by: 
\begin{equation}
T_{r}\left(x,t\right)=\left(\frac{t}{\text{ns}}\right)^{\frac{1}{5}}f^{1/4}\left(\xi_{0}x/x_{F}\left(t\right)\right)\ \text{keV},\label{eq:Tr_anal_6}
\end{equation}
\begin{equation}
T\left(x,t\right)=\left(\frac{t}{\text{ns}}\right)^{\frac{1}{5}}g^{1/4}\left(\xi_{0}x/x_{F}\left(t\right)\right)\ \text{keV}.\label{eq:Tm_anal_6}
\end{equation}
The total and absorption optical depths at the end of the simulation
are $\mathcal{T}\left(t_{\text{end}}\right)\approx97.9$ and $\mathcal{T}_{a}\left(t_{\text{end}}\right)\approx0.00026$,
so that the wave is highly optically thick and highly absorption thin,
which is not surprising since in this test the scattering opacity
is very large and the absorption opacity is very small.

In Figure \ref{fig:simulation_6} we observe an agreement between
the self-similar solution and all of the simulations, with the slight
discrepancy between the $S_{N}$ results and the other methods at
the wavefront. The $S_{N}$ results have a wave speed that is slightly
slower than the results from the other methods. Evidence from numerical
experiments indicates that this discrepancy can be mitigated by decreasing
the time step size used in the calculation. Nevertheless, there is
a limitation in this problem because there is numerical minimum for
the initial temperature imposed to keep the opacities positive, as
discussed above.


\subsection{Flux limited diffusion\label{subsec:Flux-limited-diffusion}}

\begin{figure}
\begin{centering}
\includegraphics[scale=0.55]{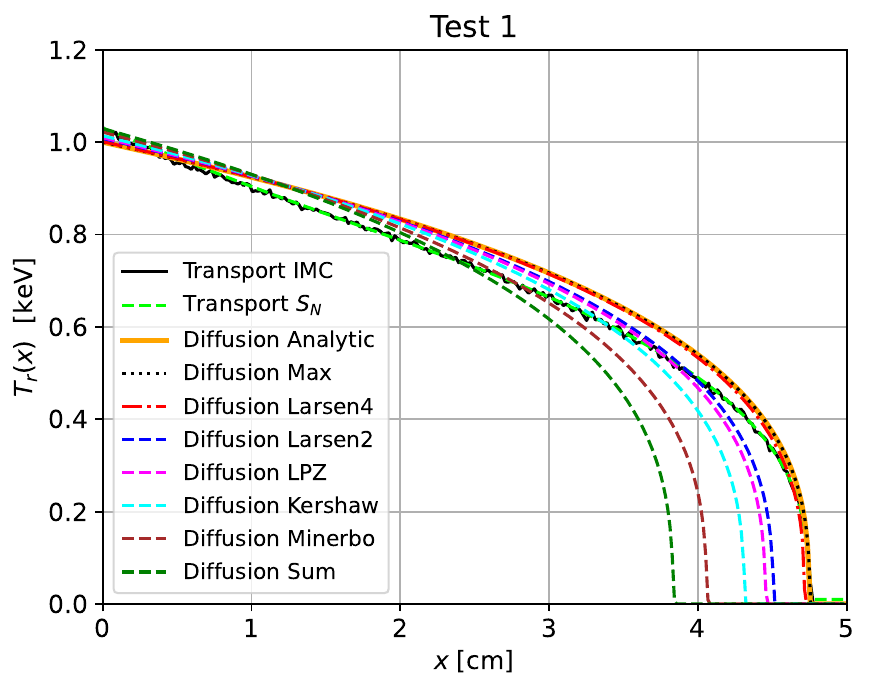} 
\par\end{centering}
\begin{centering}
\includegraphics[scale=0.55]{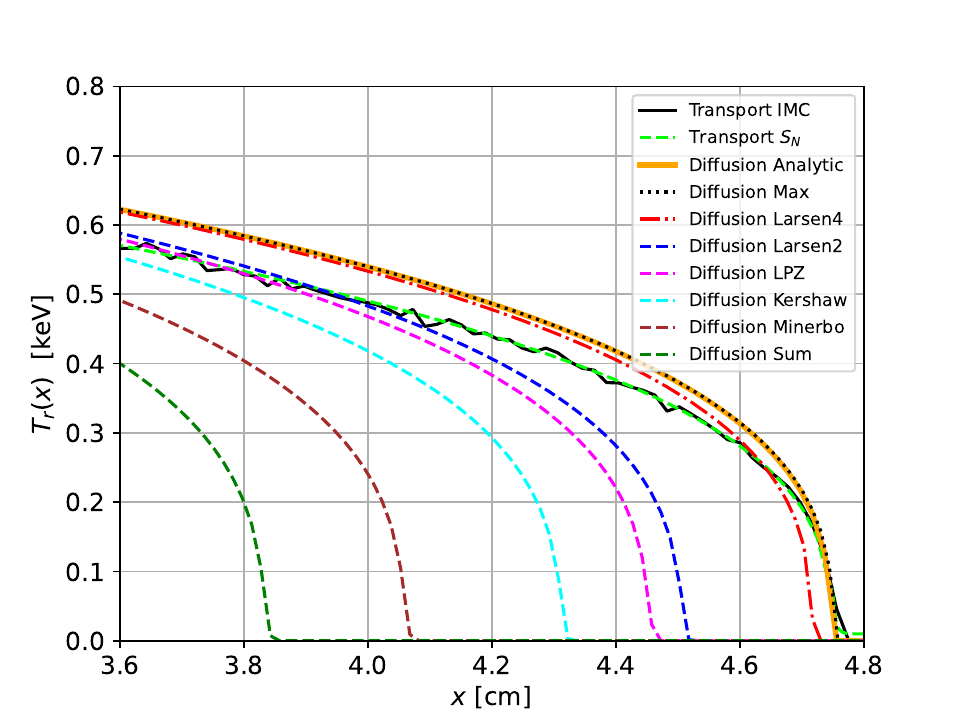} 
\par\end{centering}
\begin{centering}
\includegraphics[scale=0.55]{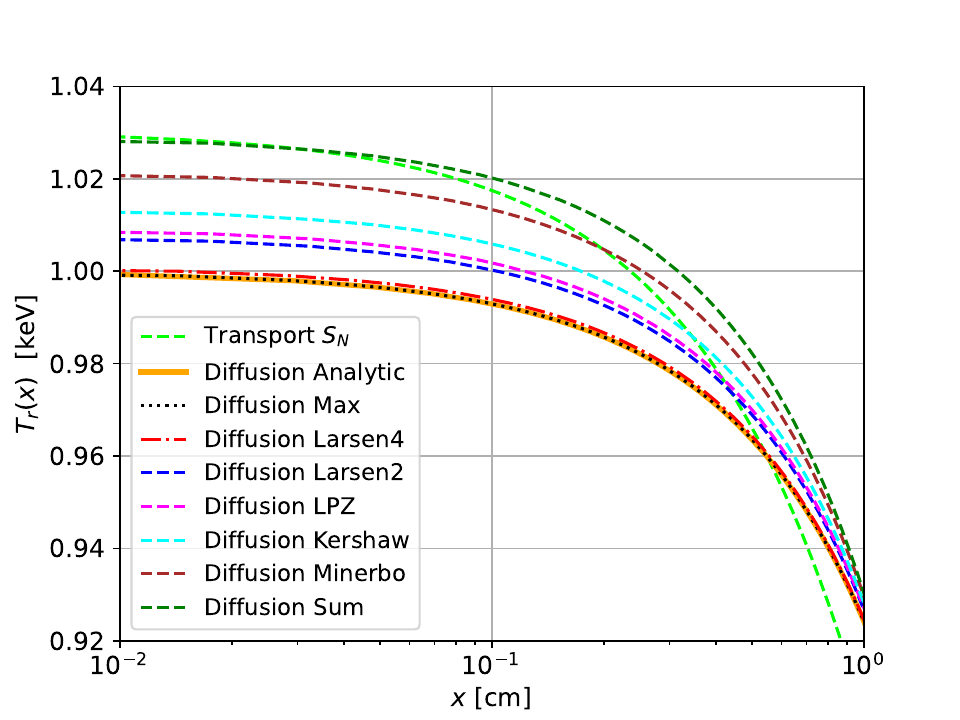} 
\par\end{centering}
\caption{A comparison of the radiation temperature profiles at the final time
of Test 1, between flux limited Diffusion simulations with various
flux limiters and the $S_{N}$ and IMC simulations (upper figure).
The middle figure is a close up view of the heat front. The bottom
figure is a close up view near the origin (excluding the noisy IMC
result). \label{fig:Tr_fl_test1}}
\end{figure}

\begin{figure}
\begin{centering}
\includegraphics[scale=0.55]{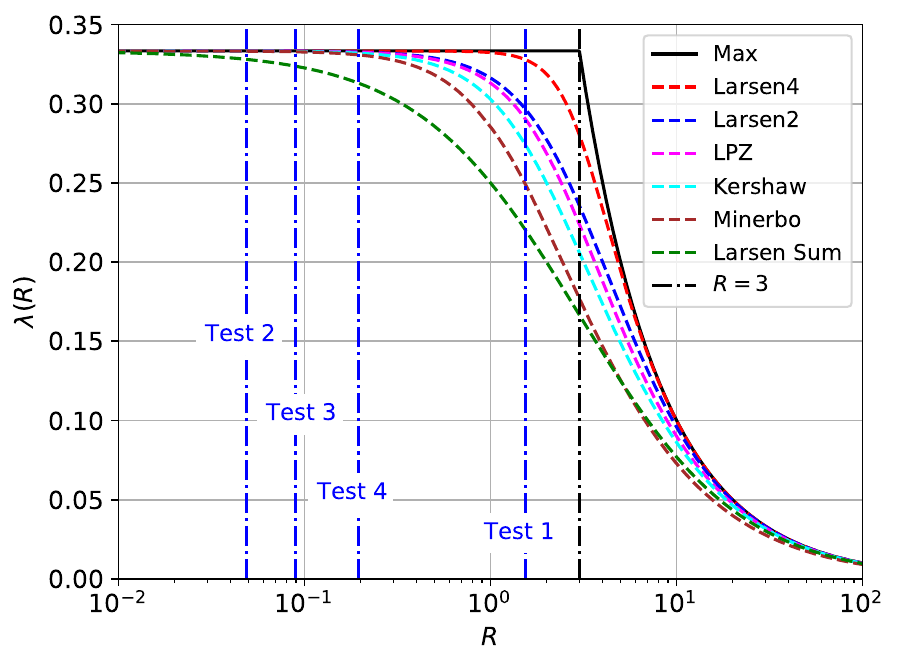} 
\par\end{centering}
\caption{Various flux limiter functions $\lambda\left(R\right)$. The vertical
dashed black line represents the free-streaming limit $R=3$. The
blue vertical lines represent the (space and time independent) values
of $R$ for Tests 1-4. \label{fig:fl_func}}
\end{figure}

In this section we discuss how a common numerical treatment of radiation
diffusion, known as flux-limited diffusion \cite{olson2000diffusion,humbird2017adjoint},
will behave on the heat waves presented above. We begin with the radiation
flux as given by Fick's law: 
\begin{equation}
F=-D\frac{\partial E}{\partial x}.
\end{equation}
This flux does not necessarily obey the free-streaming (causality)
limit that says there cannot be a bigger flux of energy than the total
amount of radiation energy present times the speed of light: 
\begin{equation}
\left|F\right|\leq cE.
\end{equation}
Flux limiters seek to rein in the flux specified by Fick's law. One
can quantify the amount of limiting required via a flux limiter parameter
given by: 
\begin{equation}
R\left(x,t\right)=\frac{3\left|F\right|}{cE}=\frac{1}{k_{t}E}\left|\frac{\partial E}{\partial x}\right|.
\end{equation}
If $R\leq3$, we have $\left|F\right|\leq cE$ and no limiting is
required. In order to obey the free-streaming limit, the flux limiter
function $\lambda\left(R\right)$ defines a flux limited diffusion
flux $F_{\text{FL}}$ in terms of a corrected diffusion coefficient
$\hat{D}$: 
\begin{equation}
F_{\text{FL}}=3\lambda\left(R\right)F=-3\lambda\left(R\right)D\frac{\partial E}{\partial x}=-\hat{D}\frac{\partial E}{\partial x}
\end{equation}
Using our analytical solution (equations \ref{eq:Ess},\ref{eq:heat_flux})
one finds an exact explicit expression: 
\begin{equation}
R\left(x,t\right)=-\sqrt{\frac{3t^{\left(\frac{\alpha}{\alpha'}-1\right)}}{ck_{0}T_{0}^{-\alpha}}}\frac{g^{\frac{\alpha}{4}}\left(\xi\right)f'\left(\xi\right)}{f\left(\xi\right)}.
\end{equation}
This expression is proportional to the inverse of the total optical
depth (see equation \ref{eq:depth_tot}): 
\begin{equation}
R\left(x,t\right)=-\frac{\xi_{0}}{\mathcal{T}\left(t\right)}\left(\frac{g\left(\xi\right)}{g_{0}}\right)^{\frac{\alpha}{4}}\frac{f'\left(\xi\right)}{f\left(\xi\right)},
\end{equation}
which is to be expected, since for optically thick waves a flux limiter
is not needed and transport results agree with diffusion results.
It is interesting to note that in the case $\alpha=\alpha'$, for
which we know the similarity profiles analytically (see section \ref{subsec:An-exact-analytic}),
the flux limiter parameter is time and space independent: 
\begin{equation}
R=\frac{12g_{0}^{\frac{\alpha}{4}}}{\alpha\xi_{0}\sqrt{3ck_{0}T_{0}^{-\alpha}}}.
\end{equation}

In Figure \ref{fig:Tr_fl_test1} we compare the results of gray diffusion
simulations using several flux limiters with the analytic diffusion
solution and transport simulations, for Test 1. The limiters compared
are the Larsen \cite{morel2000diffusion} flux-limiter and its variants,
the Minerbo flux limiter \cite{su2001variable}, the Levermore-Pomraning-Zimmerman
limiter \cite{levermore1981flux,olson2000diffusion}, and the Kershaw
limiter \cite{su2001variable}. The Larsen limiter writes the flux-limited
diffusion coefficient as: 
\begin{equation}
\hat{D}=c\left((3k_{t})^{n}+\left|\frac{\partial E}{\partial x}\right|^{n}\right)^{-1/n},
\end{equation}
which is equivalent to the flux limiter function: 
\begin{equation}
\lambda\left(R\right)=\frac{1}{3}\left(1+\left(\frac{R}{3}\right)^{n}\right)^{-1/n}
\end{equation}
The value of $n$ is a user-defined parameter and informs how the
flux-limiter transitions away from standard diffusion. The limiter
with $n=1$ is called the ``sum'' limiter, and the limit as $n\rightarrow\infty$
is known as the ``max'' limiter, which strictly enforces the free-streaming
limit. We note that all limiters mentioned above, except the max limiter,
limit the flux even for $R<3$, which is not required by the free-streaming
limit, but lead to better results in certain scenarios.

One benefit of Test 1 is that it demonstrates how different flux limiters
can give heat fronts that do not agree with either the diffusion or
transport solutions. In Figure \ref{fig:Tr_fl_test1} we see that
the Larsen with $n=4$ and the max limiters agree with the transport
results for the heat front position. None of the other flux-limited
solutions captures the behavior of the transport solutions, and result
in retarded fronts, due to the nonphysical limiting for $R<3$. Interestingly,
the Larsen limiter with $n=2$ has been used as a default setting
{\em comme il faut} \cite{brunner2005user} given that this value
preserves the asymptotic diffusion limit as $k_{t}\rightarrow\infty$
\cite{morel2000diffusion} and is smoother than the max limiter. Because
this problem has a large, but finite value for $k_{t}$, this limiter
still has an injurious effect on the heat front. This result is consistent
with Ref. \cite{cohen2021multi} in which radiative Marshak waves
experiments were analyzed via classical diffusion, flux-limited diffusion
and transport simulations.

The reason for these discrepancies can be seen by looking at how the
limiters behave as a function of $R$. In Figure \ref{fig:fl_func}
we compare the strength of the limiting, $\lambda(R)$, with the actual
values of $R$ for the Tests 1-4. We observe that, except for the
max limiter, all of the limiters affect the solution in the test problems,
with the sum limiter having the strongest effect and the Larsen with
$n=4$ limiter having the least, an order that is consistent with
the results in Figure \ref{fig:Tr_fl_test1}. We see that for Tests
2-4, which are optically thick, the effect of all flux limiters other
than the sum limiter is negligible, as expected. Moreover, the fact
that Test 1 has $R<3$, means that no flux limiting is needed in the
diffusion simulation, which results in the agreement of the \textit{heat
front} position with the transport results, shown in figure \ref{fig:simulation_1}.
However, since Test 1 defines an optically thin wave, none of the
diffusion simulations (with or without flux limiters) agree with the
\textit{shape} of the transport Marshak wave profile. We also note
that small deviations from the surface boundary temperature appear
near the origin in figure \ref{fig:Tr_fl_test1} for the flux-limited
diffusion simulations. These also result from flux-limiting, and the
use of the Marshak boundary condition (equation \ref{eq:maesh_bc_def}).
These deviations do not appear, by construction, in flux-limited diffusion
simulations which are performed with a prescribed surface temperature
boundary condition (equation \ref{eq:Tbound}), instead of the Marshak
boundary condition.

\section{Conclusion}

In this work we have developed analytic and semi-analytic self-similar
solutions to a nonlinear non-equilibrium supersonic Marshak wave problem
in the diffusion limit of radiative transfer. The solutions exist
under the assumption of a material model with power law temperature
dependent opacities and a material energy density which is proportional
to the radiation energy density, as well as a temporal power-law surface
radiation temperature drive. The solutions are a generalization of
the widely used Pomraning-Su-Olson \cite{pomraning1979non,bingjing1996benchmark}
non-equilibrium linear Marshak wave problem to the nonlinear regime.

The solutions are analyzed in detail, including a study of the LTE
limit and the non-LTE optically thick and thin limits. By inspecting
the solution near the origin, it is shown that the ratio between the
radiation and material temperatures and derivatives near the origin,
can be calculated by the root of a simple nonlinear equation. Moreover,
it is shown that for the special case for which the absorption and
total opacities have the same temperature exponents, the similarity
profiles have a simple exact analytic solution, which is essentially
a generalization of the well known Henyey LTE Marshak wave \cite{hammer2003consistent,rosen2005fundamentals},
to the non-LTE regime.

We constructed a set of six non-equilibrium Marshak wave benchmarks
for supersonic non-equilibrium radiative heat transfer. These benchmarks
were compared in detail to implicit Monte-Carlo and discrete-ordinate
radiation transport simulations as well flux limited gray diffusion
simulations. The first benchmark which is not optically thick, resulted,
as expected, in a good quantitative agreement with the diffusion simulation
and only a qualitative agreement with transport simulations. All other
benchmarks were defined to be optically thick, and resulted in a very
good agreement with transport simulations as well. All benchmarks
except Test 3 were defined to be absorption thin, resulting in a substantial
state of non-equilibrium, with a large difference between the material
and radiation temperatures. This demonstrates the usefulness of the
solution developed in this work as a non-trivial but straightforward
to implement code verification test problem for non-equilibrium radiation
diffusion and transport codes.

\subsection*{Availability of data}

The data that support the findings of this study are available from
the corresponding author upon reasonable request.

 \bibliographystyle{naturemag}
\bibliography{datab}

\pagebreak{}

\appendix

\section{Dimensional analysis\label{sec:Dimensional-analysis}}

\begin{table}[H]
\centering{}%
\begin{tabular}{|c|c|c|c|c|c|c|}
\hline 
$E$  & $U$  & $E_{0}$  & $K$  & $M$  & $x$  & $t$\tabularnewline
\hline 
\hline 
$\left[E\right]$  & $\left[E\right]$  & $\frac{\left[E\right]}{\left[\text{time}\right]^{4\tau}}$  & $\frac{\left[\text{length}\right]^{2}}{\left[E\right]^{\frac{\alpha}{4}}\left[\text{time}\right]}$  & $\frac{\left[E\right]{}^{\frac{\alpha'}{4}}}{\left[\text{time}\right]}$  & $\left[\text{length}\right]$  & $\left[\text{time}\right]$\tabularnewline
\hline 
\end{tabular}\caption{The dimensional quantities in the problem (upper line) and their dimensions
(lower line). $\left[E\right]$ denotes the dimensions of energy per
unit volume. \label{tab:The-dimensional-quantities}}
\end{table}

In this appendix, the method of dimensional analysis is employed in
order to find a self-similar ansatz for the solution of the problem
defined by the nonlinear gray diffusion model in equations \ref{eq:Eform}-\ref{eq:Uform},
with the initial and boundary conditions \ref{eq:init_cond}-\ref{eq:bc}.
The dimensional quantities which define the problem are listed in
table \ref{tab:The-dimensional-quantities}. The problem is defined
by $7$ dimensional quantities which are composed of $3$ different
units - time, length and energy density. Therefore, according the
the Pi (Buckingham) theorem of dimensional analysis \cite{buckingham1914physically,zeldovich1967physics,barenblatt1996scaling},
the problem can be expressed in terms of $7-3=4$ dimensionless variables
$\Pi_{1},\Pi_{2},\Pi_{3},\Pi_{4},$ which are given in terms of power
laws of the dimensional quantities. In order to be able to express
the solution with a single similarity independent variable $\Pi_{1}=\xi$
and two self-similar dependent variables for the radiation $\Pi_{2}=f\left(\xi\right)\propto E$
and matter $\Pi_{3}=g\left(\xi\right)\propto U$, we must require
that the remaining (fourth) dimensionless quantity $\Pi_{4}$ would
not depend on $E,U,x,t$, i.e., it should be a dimensionless independent
parameter characterizing the problem, and must have the form: 
\begin{equation}
\Pi_{4}\equiv\mathcal{A}=E_{0}^{n}K^{k}M^{m}
\end{equation}
The requirement that $\mathcal{A}$ is dimensionless results in the
following degenerate system of linear equations: 
\begin{align*}
 & -4n\tau-k-m=0\\
 & n-k\frac{\alpha}{4}+m\frac{\alpha'}{4}=0\\
 & 2k=0
\end{align*}
which has the solution: 
\begin{align*}
 & k=0\\
 & n=-m\frac{\alpha'}{4}\\
 & m\left(\alpha'\tau-1\right)=0
\end{align*}
A set of infinite non-trivial solutions exists only if: 
\begin{equation}
\tau=\frac{1}{\alpha'}
\end{equation}
In this case, by taking, without loss of generality, $m=1$, we have
$n=\frac{\alpha'}{4}$ so that: 
\begin{equation}
\mathcal{A}=E_{0}^{-\frac{\alpha'}{4}}M.
\end{equation}

The dimensionless independent similarity coordinate is written as:
\begin{equation}
\Pi_{1}=\xi=xt^{-\delta}E_{0}^{n}K^{k}.\label{eq:xsiapp}
\end{equation}
Since the radiation energy density at the system boundary, $E_{0}t^{4\tau}$,
has units of energy density, we can write the dimensionless similarity
profiles directly as: 
\begin{equation}
\Pi_{2}=\frac{E}{E_{0}t^{4\tau}}=f\left(\xi\right),
\end{equation}
\begin{equation}
\Pi_{3}=\frac{U}{E_{0}t^{4\tau}}=g\left(\xi\right).
\end{equation}
The requirement that $\xi$ is dimensionless results in the following
(non-degenerate) system of linear equations:

\begin{align*}
 & -\delta-4\tau n-k=0\\
 & n-\frac{\alpha}{4}k=0\\
 & 1+2k=0
\end{align*}
which has the solution: 
\begin{align*}
 & \delta=\frac{1}{2}\left(1+\tau\alpha\right)\\
 & n=\frac{\alpha}{4}k=-\frac{\alpha}{8}\\
 & k=-\frac{1}{2}
\end{align*}
Hence, it was shown that for the specific value of the temporal exponent
$\tau$ (equation \ref{eq:tau_ss}), there exists a self-similar solution
which is characterized by the dimensionless quantity given in equation
\ref{eq:adef}, and has the from given in equations \ref{xsi_def},\ref{eq:Ess},\ref{eq:Uss}.

We note that in order to have a self-similar solution, we had to assume
a quartic material energy-temperature relation (equation \ref{eq:eos}),
as it is clear from the derivation above and equations \ref{eq:main_eq}-\ref{eq:main_mat},
that otherwise a third dimensional constant (in addition to $K$ and
$M$) would exist. By assuming equation \ref{eq:eos}, only the dimensionless
parameter $\epsilon$ is introduced.

\section{Derivation of an exact solution\label{sec:eaxactsol}}

We look for exact solutions of the form:

\begin{equation}
f\left(\xi\right)=\left(1-\frac{\xi}{\xi_{0}}\right)^{\nu},\label{eq:f_ans_anal_app}
\end{equation}
\begin{equation}
g\left(\xi\right)=g_{0}\left(1-\frac{\xi}{\xi_{0}}\right)^{\eta}.\label{eq:g_ans_anal_app}
\end{equation}
As discussed in section \ref{subsec:The-solution-near}, we know that
$g_{0}=g\left(0\right)$ is given by the solution of equation \ref{eq:g0_eq_general}.
Plugging the ansatz \ref{eq:f_ans_anal_app}-\ref{eq:g_ans_anal_app}
into the matter equation \ref{eq:g_ode} gives:

\begin{align*}
 & -\eta\delta\xi\frac{g_{0}}{\xi_{0}}\left(1-\frac{\xi}{\xi_{0}}\right)^{\eta-1}=4\tau g_{0}\left(1-\frac{\xi}{\xi_{0}}\right)^{\eta}\\
 & -\epsilon\mathcal{A}g_{0}^{^{-\frac{\alpha'}{4}}}\left(1-\frac{\xi}{\xi_{0}}\right)^{-\frac{\alpha'}{4}\eta}\left(\left(1-\frac{\xi}{\xi_{0}}\right)^{\nu}-g_{0}\left(1-\frac{\xi}{\xi_{0}}\right)^{\eta}\right)
\end{align*}
To simplify this, we now assume $\eta=\nu$ and the equations reads:

\begin{equation}
-\nu\delta\frac{\xi}{\xi_{0}}\left(1-\frac{\xi}{\xi_{0}}\right)^{-1}=4\tau-\epsilon\mathcal{A}g_{0}^{^{-\frac{\alpha'}{4}-1}}\left(1-g_{0}\right)\left(1-\frac{\xi}{\xi_{0}}\right)^{-\frac{\alpha'}{4}\nu}
\end{equation}
This relation can be satisfied for all $\xi$ if we have a linear
denominator on second term of the right hand side, which gives: 
\begin{equation}
\nu=\frac{4}{\alpha'},
\end{equation}
and the equations reads: 
\begin{equation}
-\frac{4}{\alpha'}\delta\frac{\xi}{\xi_{0}}=4\tau\left(1-\frac{\xi}{\xi_{0}}\right)-\epsilon\mathcal{A}g_{0}^{^{-\frac{\alpha'}{4}-1}}\left(1-g_{0}\right)\label{eq:matter_eq_app}
\end{equation}
This relation is satisfied for all $\xi$ only if $-\frac{4}{\alpha'}\delta=-4\tau$,
and since $\tau=\frac{1}{\alpha'}$ (equation \ref{eq:tau_ss}) and
$\delta=\frac{1}{2}\left(1+\frac{\alpha}{\alpha'}\right)$ (equation
\ref{eq:delta_def}) we find: 
\begin{equation}
\delta=1,
\end{equation}
so that the heat front propagates at constant speed, and: 
\begin{equation}
\alpha'=\alpha,
\end{equation}
that is, the absorption opacity and total opacity have the same temperature
dependence. This is not surprising, since the solution \ref{eq:f_ans_anal_app}-\ref{eq:g_ans_anal_app}
with $\eta=\nu$ has $g'\left(0\right)/f'\left(0\right)=g_{0}$, and
therefore, from the general identity \ref{eq:gpfp_rat} can only be
valid for $\alpha=\alpha'$.

Equation \ref{eq:matter_eq_app} now reads:

\begin{equation}
0=4\tau-\epsilon\mathcal{A}g_{0}^{^{-\frac{\alpha}{4}-1}}\left(1-g_{0}\right),
\end{equation}
which results in the relation: 
\begin{equation}
g_{0}\left(1+\frac{4\tau}{\epsilon\mathcal{A}}g_{0}^{^{\frac{\alpha}{4}}}\right)=1,\label{eq:g0_rel_app}
\end{equation}
which is, as expected, identical to equation \ref{eq:g0_eq_general}
for $\alpha=\alpha'$.

We now plug the ansatz \ref{eq:f_ans_anal_app}-\ref{eq:g_ans_anal_app}
into the radiation equation \ref{eq:f_ode}: 
\begin{align*}
 & \frac{\nu\left(\nu-1\right)}{\xi_{0}^{2}}\left(1-\frac{\xi}{\xi_{0}}\right)^{\nu-2}=\\
 & \left(g_{0}\left(1-\frac{\xi}{\xi_{0}}\right)^{\nu}\right)^{-\frac{\alpha}{4}}\left[4\tau\left(1-\frac{\xi}{\xi_{0}}\right)^{\nu}+\frac{\delta\nu}{\xi_{0}}\left(1-\frac{\xi}{\xi_{0}}\right)^{\nu-1}\xi\right]\\
 & -\frac{\alpha}{4}\frac{\frac{\nu}{\xi_{0}}\left(1-\frac{\xi}{\xi_{0}}\right)^{\nu-1}\frac{\nu g_{0}}{\xi_{0}}\left(1-\frac{\xi}{\xi_{0}}\right)^{\nu-1}}{g_{0}\left(1-\frac{\xi}{\xi_{0}}\right)^{\nu}}\\
 & +\mathcal{A}\left(g_{0}\left(1-\frac{\xi}{\xi_{0}}\right)^{\nu}\right)^{-\frac{\alpha}{4}-\frac{\alpha'}{4}}\left[\left(1-\frac{\xi}{\xi_{0}}\right)^{\nu}-g_{0}\left(1-\frac{\xi}{\xi_{0}}\right)^{\nu}\right].
\end{align*}
Since $\alpha=\alpha'$, $\nu=\frac{4}{\alpha}$, $4\tau=\nu$ and
$\delta=1$, all $\xi$ dependent terms drop, and the following equation
is obtained:

\begin{align}
\nu & =g_{0}^{-\frac{\alpha}{4}}\xi_{0}^{2}+\frac{\mathcal{A}}{\nu}\xi_{0}^{2}g_{0}^{-\frac{\alpha}{2}}\left(1-g_{0}\right).
\end{align}
From equation \ref{eq:g0_rel_app} we substitute $1-g_{0}=\frac{4\tau}{\epsilon\mathcal{A}}g_{0}^{^{\frac{\alpha}{4}+1}}$,
and by solving for $\xi_{0}$ equation \ref{eq:xsi0_exact} is obtained. 
\end{document}